\documentclass[a4paper,fleqn]{cas-dc}
\usepackage[numbers]{natbib}
\usepackage{float}
\usepackage{graphicx}
\usepackage{blindtext}
\usepackage[export]{adjustbox}
\usepackage{multicol}
\usepackage[utf8]{inputenc}
\usepackage{amssymb,amsfonts}
\everymath{\displaystyle}
\usepackage[many]{tcolorbox}
\usepackage[figuresright]{rotating}
\usepackage{array}
\usepackage{kantlipsum}
\usepackage{gensymb}
\usepackage{xpatch}
\usepackage{tikz}
\usepackage{lipsum}
\usepackage{multirow} 
\usepackage{lineno}
\usepackage{amsmath}
\usepackage{textcomp}
\usepackage{amsbsy,enumerate}
\usepackage{chemformula}
\usepackage{balance}
\usepackage{geometry}
\usepackage{ragged2e}
\usepackage{setspace}
\usepackage{etoolbox}
\usepackage{siunitx}
\usepackage{mathtools}
\usepackage{xcolor}
\usepackage{fullpage,anysize,esint,relsize, commath}
\usepackage{fancyhdr} 
\usepackage{hyperref}
\usepackage{enumitem}
\usepackage{pifont}
\usepackage{listings}
\usepackage{cprotect}
\usepackage{multiaudience}
\usepackage{abbrevs}
\usepackage{isomath}
\usepackage{graphics}
\usepackage{stfloats}

\DeclareMathAlphabet{\mathpzc}{OT1}{pzc}{m}{it}
\DeclareMathAlphabet{\mathbbm}{U}{bbm}{m}{n}

\def\tsc#1{\csdef{#1}{\textsc{\lowercase{#1}}\xspace}}
\tsc{WGM}
\tsc{QE}
\tsc{EP}
\tsc{PMS}
\tsc{BEC}
\tsc{DE}
\begin{document}
\let\WriteBookmarks\relax
\def\floatpagepagefraction{1}
\def\textpagefraction{.001}

\title [mode = title]{A Survey of Malware Detection Using Deep Learning}
\author[1]{Ahmed Bensaoud}
\cormark[1]
\ead{abensaou@uccs.edu}
\author[1]{Jugal Kalita}
\ead{jkalita@uccs.edu}
\address[1]{Deptarment of Computer Science, University of Colorado Colorado Springs, CO, USA}
\author[1]{Mahmoud Bensaoud}
\ead{mbensao2@uccs.edu}
%\address[2]{Air Academy High School Colorado Springs, CO, USA}
\cortext[cor1]{Corresponding author:}
 
\begin{abstract}
The problem of malicious software (malware) detection and classification is a complex task, and there is no perfect approach. There is still a lot of work to be done. Unlike most other research areas,  standard benchmarks are difficult to find for malware detection. This paper aims to investigate recent advances in malware detection on MacOS, Windows, iOS, Android, and Linux using deep learning (DL) by investigating DL in text and image classification, the use of pre-trained and multi-task learning models for malware detection approaches to obtain high accuracy and which the best approach if we have a  standard benchmark dataset. We discuss the issues and the challenges in malware detection using DL classifiers by reviewing the effectiveness of these DL classifiers and their inability to explain their decisions and actions to DL developers presenting the need to use Explainable Machine Learning (XAI) or Interpretable Machine Learning (IML) programs. Additionally, we discuss the impact of adversarial attacks on deep learning models, negatively affecting their generalization capabilities and resulting in poor performance on unseen data. We believe there is a need to train and test the effectiveness and efficiency of the current state-of-the-art deep learning models on different malware datasets. We examine eight popular DL approaches on various datasets. This survey will help researchers develop a general understanding of malware recognition using deep learning. 
\end{abstract}

\begin{keywords}
Malware Detection \sep Multi-task Learning \sep Malware Image \sep Generative Adversarial Networks \sep Mobile Malware \sep Convolutional Neural Network
\end{keywords}
\maketitle
\section{Introduction}
Operating systems such as Windows, Android, Linux, and MacOS are updated every few weeks to protect against critical vulnerabilities. On the other hand, malware authors are also always looking for new ways to finesse their malicious code to overwhelm the new operating system updates. Every operating system is vulnerable. In addition, since operating systems run on desktops and servers, and even on routers, security cameras, drones and other devices, the biggest problem is diversity of systems to protect because all these devices are very different. 

\begin{figure*}[ht]
\centering
\includegraphics[height=80mm,width=160mm]{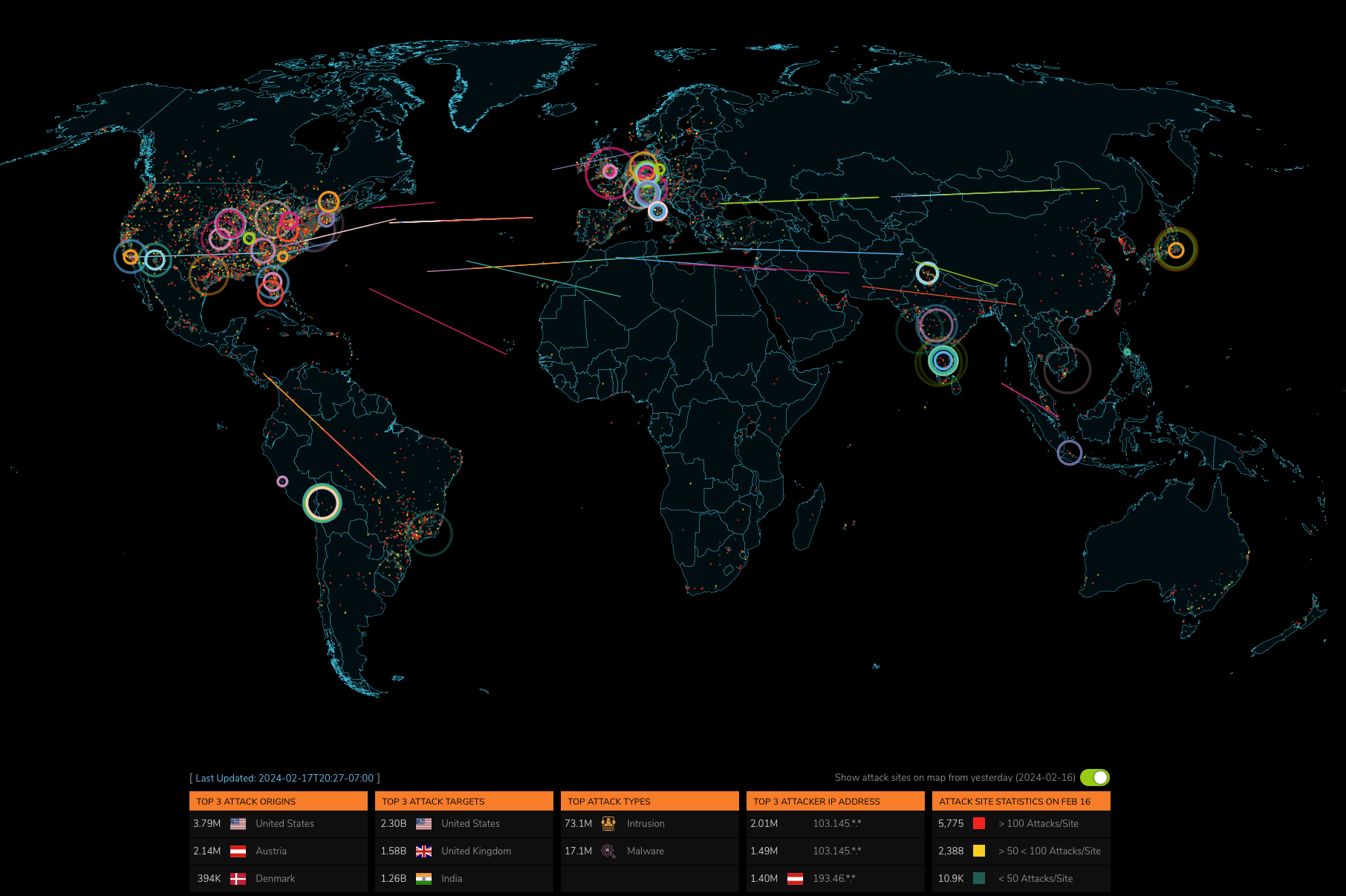}
  \caption{Worldwide attacks}
  \label{mylabel1}
\end{figure*}

Most every day, there is a new story about malicious software in the news. For example, in Oct 2022,  cyberattacks coming from a Russia-based hacker group known as Killnet targeted the government services of the state of Colorado, Alabama, Alaska, Delaware, Connecticut, Florida, Mississippi, and Kansas websites\footnote{https://www.nbcnews.com/tech/security/colorado-state-websites-struggle-russian-hackers-vow-attack-rcna51012}.  Again in 2022, hackers working on behalf of the Chinese government stole \$20 million from covid relief benefits\footnote{https://www.nbcnews.com/tech/security/china-hacked-least-six-us-state-governments-report-says-rcna19255}. 
The increase in the vulnerability of sensitive data due to cyber-attacks, cyber-threats, cyber-crimes, and malware needs to be countered. In 2023, Fig. ~\ref{mylabel1} shows countries that have been attacked by malware and the top origins of these malware \footnote{https://attackmap.sonicwall.com/live-attack-map}.

Researchers have used deep learning to classify malware samples since it generalizes well to unseen data. Our survey focuses on static, dynamic and hybrid malware detection methods in Windows, Android, Linux, MacOS, and iOS. We describe the strengths and weaknesses of deep learning models for malware detection. 
Most recent research uses deep neural networks (DNNs) for malware classification and achieves high success. State-of-the-art DNN models have been developed against modern malware such as Zeus, Fleeceware, RaaS, Mount Locker, REvil, LockBit, Cryptesla, Snugy, and Shlayer.\\
The contributions of this paper are as follows:

\begin{itemize}
\item It gives the big picture of how hackers attack (Sections 2,3,4,5).
\item It presents how to generate images form malware files (Section 6).
\item It discusses deep learning models for malware image classification (Section 7).
\item It describes feature reduction that can improve performance (Section 8).
\item It discusses transfer learning approaches in the classification of malware and what needs to improve for better performance (Section 9).
\item It reviews the use of natural language processing in malware classification (Section 10). 
\item It presents the deep learning models for cryptographer ransomware (Section 11).
\item  It shows how we know if we can trust the results of a DL model using Explainable Artificial Intelligence, XAI (Section 12).
\item  It discusses significant challenge for the reliability and security pozed by adversarial attacks on deep learning models (Section 13).
\end{itemize}

The rest of this paper, we discuss avenues for future research and we examine the Efficientnet B0, B1, B2, B3, B4, B5, B6, and B7 models on malware images datasets for classification.
\section{Mechanics of Malware Attacts}
The hacker has one goal, which is to get malware installed onto a victim’s computer. Because most computers are protected by some type of firewall, direct attacks are difficult to impossible to perform. Therefore, attackers attempt to trick the computer into running the malicious code. The most common way to do this is by using documents or executable files. For instance, a hacker may send an email or a phish to the victim with a malicious document attachment or a link to a website where the malicious document is located. Once the victim opens the document, embedded exploits or scripts run and download or extract more malware. This is the real malware the hacker wants to run on the victim’s system and is often something like a backdoor or ransomware. However, malicious documents are usually not the final piece of malware in an attack, but are one of the compromised vectors used by the hacker to get on the system. As an example, below we discuss how a PDF document can be used to initiate an attack.   

\subsection{PDF and Document Files}
When analyzing PDF, we find three things: \textbf{Object}, which is the structure of the PDF, \textbf{Keywords}  which control how the PDF works, and \textbf{Data} stored or encoded within a PDF.
\begin{itemize}
\item \textbf{Objects} are the building blocks of a PDFs. Every PDF starts with a Header which needs to be present in the first 1024 bytes of the documents. Some hackers take advantage of this by putting unrelated data within the first 1024 bytes. This is a very simple technique to try to avoid signature-based detection. PDFs are composed of objects; each section has specific data within the document or performs a specific function. Each object starts with two numbers, followed by the keyword obj, and ends with endobj. There are many kinds of objects, such as font objects, image objects, and even objects that contain metadata.

\item There are many \textbf{keywords} that begin with a \slash and describe how the PDF works. Some of the keywords related to malicious activity include \textsl{\slash{OpenAction}}, or its abbreviation \textsl{\slash{AA}}, both of which indicate an automatic action to be performed when the document is viewed\footnote{https://blog.didierstevens.com/programs/pdf-tools/}. This keyword points to another object that automatically gets opened or executed when the PDF is opened. Malicious PDFs have \textsl{\slash{OpenAction}} pointing to some malicious JavaScript, or an object containing an export; whenever one opens the document, the system is automatically compromised. \textsl{\slash{JavaScript}} or \textsl{\slash{JS}} keyword indicate the presence of JavaScript code. Malicious PDFs usually contain malicious JavaScript to launch an exploit or download additional malware. Some objects can be referred to as \textsl{\slash{Name}} instead of their number. Some PDFs have the ability to have files embedded with keyword \textsl{\slash{EmbeddedFile}}, \textsl{\slash{URL}} or \textsl{\slash{SubmitForm}}. \textsl{\slash{URL}} is accessed or downloaded when the object is loaded. 
\item 
PDFs can encode \textbf{data} in multiple ways, which is very flexible and can store data in a number of ways. Hackers can encode and hide their data. For example, names are case sensitive, but can be fully or partially hex encoded. More precisely, the \# sign followed by two hex characters represents hex encoded data. Data also can be octal encoded or represented by their base eight number. The octal encoded character has a $\backslash$ followed by three digits between 0 and 7. However, the hackers can mix hex, octal, and ASCII data all together, which makes it possible to hide data such as JavaScript code or URLs. 
\end{itemize}

\begin{figure} [ht]
	\includegraphics[width=\linewidth, frame]{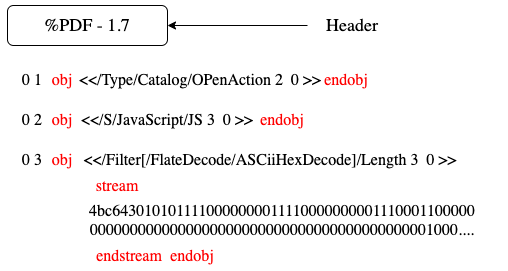}
	\caption{PDF format example.}
\label{mylabel2}
\end{figure}

\begin{figure} [ht]
	\includegraphics[width=\linewidth, frame]{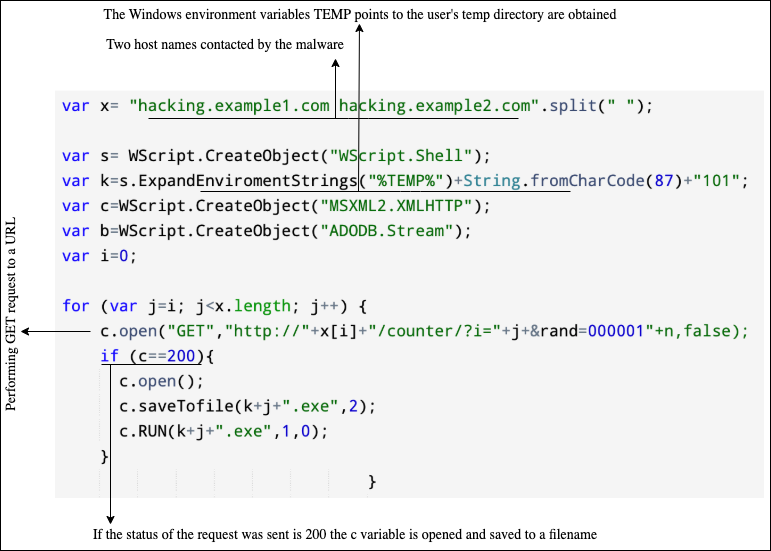}
	\caption{Malicious JavaScript code}
 \label{mylabel3}
\end{figure}

The names and strings can be encoded, but data streams can be modified and encoded further using filters. \textbf{Filters} are algorithms that are applied to the data to encode or compress within the PDF. There are multiple filters that can be used in PDFs, such as \textsl{\slash{ASCiiHexDecode}}, Hex encoding of characters; \textsl{\slash{LZWDecode}}, LZW compression algorithm; \textsl{\slash{FlateDecode}}, Zlib compression; \textsl{\slash{ASCii85Decode}}, ASCII base-85 representation; and \textsl{\slash{Crypt}}, various encryption algorithms. For example, in Fig. ~\ref{mylabel2}, we have a PDF document with three objects. Object 1 is a catalog that has OpenAction and is referring to version 0 of object 2, which means as soon as the document is opened, Object 2 will be run. Object 2 contains a JavaScript keyword, but we do not see any JavaScript code in this object because the JavaScript keyword refers to another object which is Object 3. Object 3 is a stream object as indicated by the stream keyword and has been ASCiiHex encoded and compressed with the Zlib compression algorithm. However, we have been able to determine that as soon as the PDF opens, JavaScript will be executed, and we do not know what the JavaScript's goal is. If this is a malicious PDF, it can cause problems. In Fig.  ~\ref{mylabel3}, the JavaScript code references the two hosts' names, performs an HTTP GET request to each, saves an executable file, and finally runs it.

Is malicious JavaScript used only in documents? The answer is everywhere. Malicious JavaScript is used in web pages that are created by web attack kits that perform drive-by downloads. The user opens the website that has been compromised or loads a malicious ad,  which then loads malicious JavaScript. Without JavaScript, it is difficult for hackers to get their exploit to work.

\begin{figure} [ht]
	\includegraphics[width=\linewidth, frame]{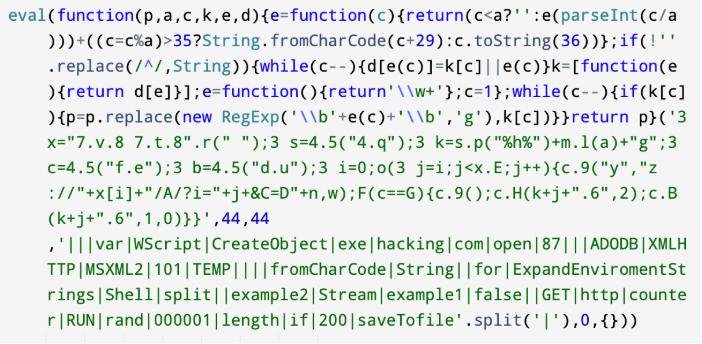}
	\caption{Obfuscated malicious JavaScript code.}
 \label{mylabel4}
\end{figure}

Most hackers try to hide what their script is doing using obfuscation techniques. Most techniques used to obfuscate script can be broken down into four different categories. How  the \textbf{format} of a program is obfuscated is shown in Fig.  ~\ref{mylabel4}; approaches include adding \textbf{extra lines of code}, \textbf{obfuscating the data}, and \textbf{substituting} variable names. 

\section{Nature of Malware Code}
The nature of malware code encompasses various characteristics and behaviors that define its purpose and functionality. Malware, short for malicious software, refers to any code or program designed with malicious intent to compromise systems, steal information, or disrupt normal operations. The nature of malware code can vary depending on its specific type and objectives, but some common attributes include: 
\subsection{Obfuscation} 
Obfuscation is an attempt by an author of a piece of code to obscure the meaning, to make something unclear, or make it very difficult to analyze. It may use encryption or compression to hide its true intentions or to evade signature-based detection by security software.
\subsection{Payload Delivery}
Malware code typically carries a payload, which is the malicious action it intends to execute. This can range from stealing sensitive information (e.g., financial data, login credentials ) to launching distributed denial-of-service (DDoS) attacks, encrypting files for ransom (ransomware), or providing backdoor access for remote control.
\subsection {Command and Control (C\&C)}
Many malware strains establish communication channels with remote servers or command-and-control infrastructure. This allows attackers to remotely control and manage the infected systems, update the malware, and receive stolen data.
\subsection{Self-Replication}
Many malware strains possess the ability to self-replicate, allowing them to spread across networks, devices, or files. This replication can occur through various means, such as attaching to exploiting vulnerabilities, legitimate files, or utilizing network resources.
\subsection{Exploitation}
Malware leverages vulnerabilities and weaknesses in software, operating systems, or user behavior to gain unauthorized access or control. It can exploit security flaws, network vulnerabilities, or social engineering techniques to compromise systems and execute malicious actions.
\subsection{Polymorphism}
Some malware utilizes polymorphic or metamorphic techniques to dynamically change its code structure or appearance while preserving its functionality. This makes it more challenging for antivirus software to detect and block.

\subsection{Ransomware}
A ransomware usually combines cryptography with malware. \textbf{How does it work?} The hacker sends the file to an unknowing victim. When the victim opens the file, it executes the malware's payload and encrypts victim data such as photos, documents, multimedia, files, and even confidential records. The hacker offten forces the victim to pay in cryptocurrency, in most cases Bitcoin.

Ransomware has worm-like properties and has names such as \textbf{WannaCrypt}, \textbf{WanaCrypt0r}, \textbf{WCRY}, \textbf{WanaDecrypt0r}, and \textbf{WCrypt}.
Each encrypted file is locked by a different key and encrypted with the RSA algorithm, which makes the file unaccessible to the owner who does not have the keys. The WannaCry virus can encrypt a large number of file types. An exhaustive list is given in Appendix A. \\

\begin{figure} [ht]
	\includegraphics[width=\linewidth, frame]{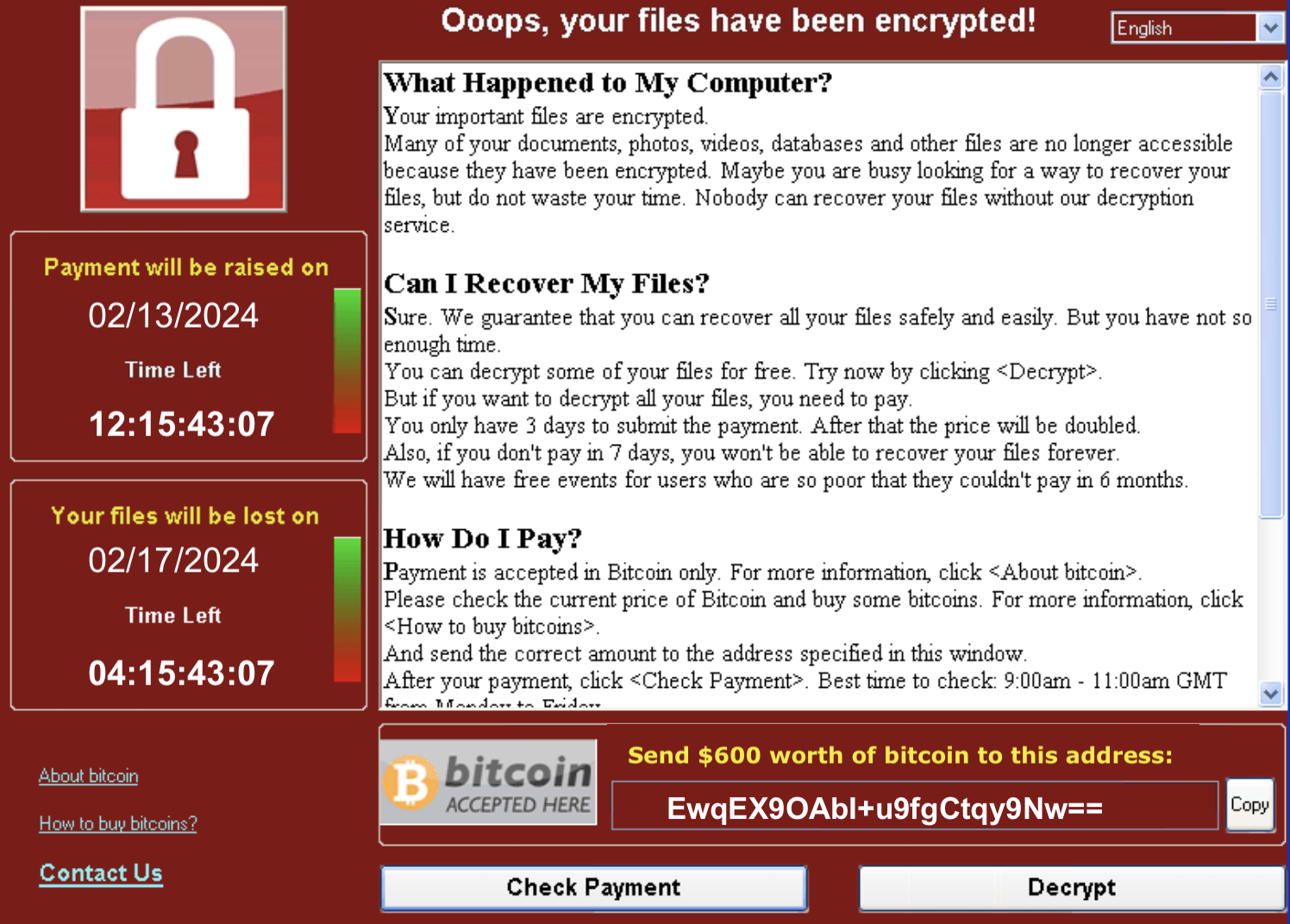}
	\caption{Oops, your files have been encrypted!}
 \label{mylabel5}
\end{figure}

The ransomware replaces the desktop wallpaper with the ransom note file by modifying Windows registry. It holds all files hostage to demand ransom payments of \$300 and later \$600 in the Bitcoin cryptocurrency as shown in Fig.  ~\ref{mylabel5}. 

As an example, on May 11, 2022, Costa Rica’s newly elected president had to declare a state of national emergency due to a ransomware attack carried out by the Conti ransomware gang. They requested \$10 million, but the demand changed to \$20 million after Costa Rica refused to pay\footnote{https://securityintelligence.com/news/costa-rica-state-emergency-ransomware/}. As another example, in october 2022, ransomware gang accessed data on 270,000 patients from Louisiana hospital system \footnote{https://www.cnn.com/2022/12/28/politics/hackers-access-data-louisiana-hospital-system-ransomware/index.html}.

Understanding the nature of malware code is crucial for developing effective defense mechanisms and mitigating its impact. It enables security professionals to develop robust detection methods, implement security best practices, and respond promptly to evolving threats.

\section{Overview \& Malware Detection}
Malware detection methods are divided into three types: static, dynamic, and hybrid 
\cite{damodaran2017comparison}. Static methods inspect an executable file without running it, while dynamic methods must run the executable file and analyze its behaviors inside a controlled environment. In hybrid methods, the information is collected regarding malware from static as well as dynamic analysis.

Some security researchers use static features by decompiling the target file. \citet{Nitin22} proposed a fuzzy-import hashing technique based on static analysis for malware detection. \citet{mohamad2021static} proposed machine learning classifiers based on permission-based features for static analysis to detect Android malware. 

Compared to static analysis, dynamic analysis includes system dynamic behavior monitoring, snapshot, debugging, etc. \citet{Kim8622568} presented a new encoding technique for dynamic features to identify anomalous events using Convolutional Neural Networks (CNNs).

Security researchers have also extracted combined features from different parts of malware files. 
\citet{BAI2021107639} extracted features from static and dynamic analysis of Android apps and applied a deep learning technique. \citet{Chaulagain1} presented a deep learning-based hybrid analysis technique by collecting different artifacts during static and dynamic analysis to train the deep learning models.

\section{Data for Malware Detection}
Numerous system logs of activities of machines such as phones, tablets, laptops, and other devices are generated by the operating system and other infrastructure software. The data are created and stored on the local device and sent to remote servers. Analyzing log data, we can not only detect breaches or suspicious activity, but we can track behavior through the network. Log data allow us to track security events, troubleshoot the infrastructure, and optimize the environment and the machines. Log data can take many different forms like syslog, authentication logs, local security event logs, network asset logs, and system logs. One of goals in malware detection is to be able to read, search, and analyze the data efficiently and effectively. 

Table 1 contains some information that is useful from syslog and windows logs. Both kinds of logs have many components in different format that helps us in the investigation.

\begin{table*}
    \caption{Syslog and Windows log}
    \centering
    \begin{tabular}{ll}
        \toprule
        \textbf{Syslog} & \textbf{Windows Logs} \\
        \midrule
        IETF standard & Event log \\
        Timestamp & Contains source, event ID, and log level \\
        Standard for network equipment logging & Logs Application, security, network events from a machine or server \\
        \begin{tabular}[c]{@{}l@{}}Device-ID, severity level, message number,\\ message text\end{tabular} & Timestamp, user, computer, and process ID \\
        \begin{tabular}[c]{@{}l@{}}Can be customized on network equipment \\for different events and severity levels\end{tabular} & Used in most enterprise environments running Windows \\
        \bottomrule
    \end{tabular}
\end{table*}
\section{Generating Malware Images for Deep Learning }

Several tools can visualize and edit a binary file in hexadecimal or ASCII formats such as IDA Pro\footnote{https://hex-rays.com/ida-pro}, x32/x64 Debugger\footnote{https://x64dbg.com/\#start}, HxD\footnote{https://mh-nexus.de/en/hxd}, PE-bear\footnote{https://hshrzd.wordpress.com/pe-bear}, Yara\footnote{https://yara.readthedocs.io/en/stable}, Fiddler\footnote{https://www.telerik.com/purchase/fiddler}, Metadata\footnote{https://www.malwarebytes.com/glossary/metadata}, XOR analysis\footnote{https://eternal-todo.com/var/scripts/xorbruteforcer}, and Embedded strings\footnote{https://virustotal.github.io/yara/}.

Malware file or code can be used to generate an image by converting the binary, octal, hexadecimal or decimal into a two dimensional matrix of pixels. The image can be grayscale or RGB. In greyscale, pixels are black and white values in the range [0-255] where 0 represents black, and 255 represents white. \\

\begin{figure} [ht]
	\includegraphics[width=\linewidth, frame]{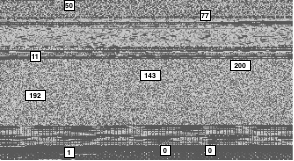}
	\caption{Malware feature representation in grayscale image}
 \label{mylabel6}
\end{figure}

\textbf{Gray image feature}: The machine stores images in a matrix of numbers. These numbers, or the pixel values, denote the intensity or brightness of the pixel. Smaller numbers (close to zero) represent black, and larger numbers (closer to 255) denote white (see Fig.  ~\ref{mylabel6}).
\\

\begin{figure} [ht]
	\includegraphics[width=\linewidth, frame]{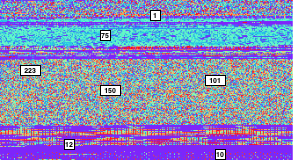}
	\caption{Malware feature representation in RGB image}
 \label{mylabel7}
\end{figure}

\textbf{RGB images}: There are three matrices or channels (Red, Green, Blue), where each matrix has values between $0-255$. These three colors are combined together in various ways to represent one of 16,777,216 possible colors (see Fig.  ~\ref{mylabel7}).\\

Malware can be converted to images in different ways. \citet{YUAN2020101740} converted malware binaries into Markov images by computing transfer probability of bytes where each pixel is generated by equation~\ref{eq:equation1}:

\begin{equation}\label{eq:equation1}
  p_{m,n}=P(n|m)=\frac{f(m,n)}{ \displaystyle\sum_{n=0}^{255} f(m,n)} \quad m, n \in \{0, 1, ..., 255\}.\\  
\end{equation}

\citet{mohammed2021malware} used a vector of 16-bit signed hexadecimal numbers to represent a $256\times256$ image. Then, they computed bi-gram frequency counts which they used as pixel intensity values. Full-frame Discrete Cosine Transform (DCT) \cite{khayam2003discrete} was computed to de-sparsify, and the bigram-DCT was used to represent the output image. \citet{Euh9057637} proposed Window Entropy Map (WEM) to visualize malware as an image. They calculated the entropy for each byte to measure the degree of uncertainty. \citet{ NI2018871} converted  malware code into gray images using SimHash \cite{charikar2002similarity} and then encoded them. They mapped SimHash values to pixels and then converted them to grayscale images.

\section{Image Classification for Malware Detection}

Deep learning can solve diverse "vision" problems, including malware image classification tasks. Deep learning can extract features automatically obviating manual feature extraction. The content of the malware executable file is first converted into a digital image.  \citet{nataraj2011malware} visualized the byte codes of samples from 25 malware families as grayscale images. Several visualization techniques have been used for malware classification. The basic idea used in these methods is to explore the distinguishing patterns in malware images. In addition, the visualization techniques help find the correlations among different malware families. Some existing approaches generate grayscale images and others generate RGB images. Most existing approaches use global features to generate malware image.

\citet{YUAN2020101740} proposed a method based on Markov images according to the byte transmission probability matrix. They used a CNN to classify Markov malware images without scaling. \citet{narayanan2020ensemble} proposed an ensemble approach using RNN and CNN architectures for malware image classification. Images were generated from assembly compiled files and classified using CNNs.  \citet{zhu2021task} proposed a Task-Aware Meta Learning-based Siamese Neural Network to classify obfuscated malware images. Their model showed high effectiveness on unique malware signature detection to classify obfuscated malware. \citet{chauhan2022classification} visualized malware files in different color modes, RGB, HSV, greyscale, and BGR. They used a support vector machine (SVM) to classify these malware images, with accuracy of 96\% in all modes. \citet{DAREM2021314} designed a semi-supervised method based on malware image and feature engineering for obfuscated malware detection.  The model achieved 99.12\% accuracy on obfuscated malware detection. \citet{asam2021detection} proposed two malware image classification approaches called Deep Feature Space-based Malware classification (DFS-MC) and Deep Boosted Feature Space-based Malware classification (DBFS-MC). The approach achieved a good accuracy of 98.61\% on the MalImg malware dataset.

\citet{XIAO2021102420} presented a visualization method called Colored Label boxes (CoLab) to specify each section in a PE file and convert it to malware image. The authors built a composed CoLab image,cand used VGG16, and Support vector machine for classification. The model was applied on two datasets, VX-Heaven\footnote{https://archive.org/download/vxheavens-2010-05-18} and BIG-2015, with 96.59\% and 98.94\% average accuracies, respectively. A comparison of reviewed malware images classification is discussed in Table 2.

\begin{table*}
    \caption{Comparative performance summary of Transfer Learning models for malware image classification.}
    \centering
    \begin{tabular}{lllllc}
        \toprule
        \textbf{Reference} & \textbf{Features} & \textbf{Model} & \textbf{Files} & \textbf{Accuracy} & \textbf{Dataset} \\
        \midrule
        \citet{CAYIR2021102133} & gray-scale images & CapsNet & PE & 98.63\% & Malimg \\
        \citet{CAYIR2021102133} & gray-scale images & RCNF & PE & 98.72\% & Malimg \\
        \citet{GoJin} & gray-scale images & ResNeXt & PE & 98.32\% & Malimg \\
        \citet{bensaoud2020classifying} & gray-scale images & Inception V3 & PE & 99.24\% & Malimg \\
        \citet{Shafai} & gray-scale images & VGG16 & PE & 99.97\% & Malimg \\
        \citet{hemalatha2021efficient} & gray-scale images & DenseNet & PE & 98.23\% & Malimg \\
        \citet{hemalatha2021efficient} & gray-scale images & DenseNet & PE & 98.46\% & BIG 2015 \\
        \citet{Lo8763852} & gray-scale images & Xception & PE & 99.03\% & Malimg \\
        \citet{Lo8763852} & gray-scale images & Xception & PE & 99.17\% & BIG 2015 \\
        \bottomrule
    \end{tabular}
\end{table*}

\section{Feature Reduction for Efficient Malware Detection}
Feature Reduction reduces the number of variables or features in the representation of a data example. Approaches to feature reduction can be divided into two subcategories called a) Feature Selection which includes methods such as Wrappers, Filters, and Embedded, and b) Feature Extraction, which includes methods such as Principal Components Analysis \cite{barath2016pattern}. \textbf{How does Feature Reduction improve performance?} It does by reducing the number of features that are considered for analysis.

In feature extraction, we start with $n$ features $x_1,x_2,x_3\\,....,x_n$, which we map to a lower dimensional space to get the new features $z_1,z_2,z_3,....,z_m$ where $m<n$. Each of the new features is usually linear a combination of the original feature set $x_1,x_2,x_3,....,x_n$. Thus, each new feature is obtained as a function F(X) of the original feature set X. This makes a projection of a higher dimensional feature space to a lower dimensional feature space, so that the smaller dimensional feature set may lead to better classification or faster classification (see equation~\ref{eq:equation2}).

\begin{equation}\label{eq:equation2}
\begin{bmatrix}
    z_{1} 
    \hdots 
    z_{m}
\end{bmatrix}^\intercal
\
=F\left(
\begin{bmatrix}
    x_{1} 
    \hdots 
    x_{n} 
\end{bmatrix}^\intercal \right)
\end{equation}

In feature selection, we choose a subset of the features, in contrast to feature extraction where we map the original features to a lower dimensional space. The smaller dimensional feature set can help produce better as well as faster classification. To do that, we need to find a projection matrix $W \ni   \bar{Z}= W^T \bar{X}$. 
We expect from such a projection that the new features are uncorrelated and cannot be reduced further and are non redundant. 
Next, we need features to have large variance: Why? Because if a feature takes similar values for all the instances, that feature cannot be used as a discriminator.  \\

Feature extraction methods such as a Principal Component Analysis (PCA) \cite{barath2016pattern}, GIST \cite{oliva2001modeling}, Hu Moments \cite{hu1962visual}, Color Histogram \cite{swain1991color}, Haralick texture \cite{lin2004comparison}, Discrete Wavelet Transform (DWT) \cite{kancherla2016packer}, Independent Component Analysis (ICA) \cite{herault}, Linear discriminant analysis (LDA) \cite{Fan5893948}, Oriented Fast and Rotated BRIEF (ORB) \cite{Rublee6126544}, Speeded Up Robust Feature (SURF) \cite{bay2006surf}, Scale Invariant Feature Transform (SIFT) \cite{Lowe790410}, Dense Scale Invariant Feature Transform (D-SIFT) \cite{Lowe790410}, Local Binary Patterns (LBPs) \cite{ojala1996comparative}, KAZE \cite{alcantarilla2012kaze} have been combined with machine learning including deep learning. These methods successfully filter the characteristics of malware files.\\

\citet{AZAD202254} proposed a method named DEEPSEL (Deep Feature Selection) to identify malicious codes of 39 unique malware families. Their model achieved an accuracy of 83.6\% and an F-measure of 82.5\%. \citet{Tobiyama7552276} proposed feature extraction based on system calls. Recurrent Neural Network was used to extract features and Convolutional Neural Network to classify these features.\\  

\section{Deep Transfer Leaning models for Malware detection}
Transfer learning takes place if we have a source model which has some pre-trained knowledge and this knowledge is needed as the foundation to build a new model \cite{YE2021107617}. For example, using a very large pre-trained convolutional neural network usually involves saving a network that was previously trained on some large dataset, typically on a large-scale image classification task, using a dataset like ImageNet \cite{russakovsky2015imagenet}. After training a network on the ImageNet dataset, we can re-purpose this trained network. Research papers have discussed applying these pre-trained networks to malware image datasets \cite{vasan2020imcfn,rezende2017malicious,bhodia2019transfer,qiao2020malware} that are generated form PE and APK malware files, which are quite different from each other.

Malware image datasets are very different from ImageNet, which is normally used to pre-train the model. The ImageNet dataset and a malware image dataset represent visually completely different images. However, pre-trained still seems to help. Training a machine learning algorithm on large datasets can be done in two ways, as discussed below. 

\subsection{Using feature extraction} 
\begin{figure} [ht]
	\includegraphics[width=\linewidth, frame]{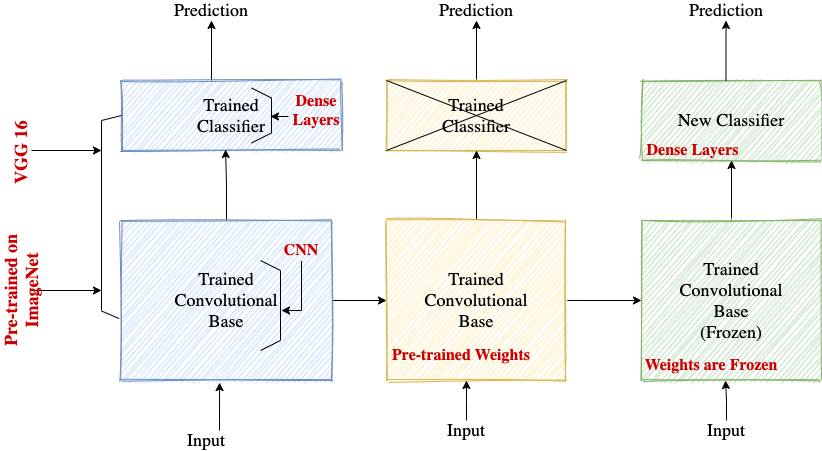}
	\caption{Feature Extraction for Transfer Learning}
 \label{mylabel8}
\end{figure}

Feature extraction discussed earlier is a practical and common, and low resource-intensive way of using pre-trained networks. It takes the convolutional base of a previously trained network and runs the malware data through it, and then trains a new classifier on top of the output. As shown in Fig. ~\ref{mylabel8}, we can choose a network such as VGG16 \cite{simonyan2014very} that has been trained on ImageNet, as an example. The input fed at the bottom, goes up to the trained convolutional base, representing the CNN region of the VGG16. The trained classifier resides in the dense region and the prediction is made by this dense region at the end. Usually, we have $1000$ neurons at the end to predict the actual ImageNet classes. We take this ImageNet trained model as base, and remove the classifier layer, keeping the convolutional layers of the pre-trained model, along with their weights. In the next step, we attach a new classifier that has new dense layers for malware classification on top. The weights of the base are frozen, which means that the malware input passes through convolutional layers which have their prior weights, during training. However, all dense layers are randomly initialized, and the interconnection weights for these layers are learned during the new training process for detecting malware.

Why remove the original dense layers? What has been observed is that the representations learned by the convolutional base are generic and therefore reusable for a variety of tasks. 

\subsection{Using fine tuning}

\begin{figure} [ht]
	\includegraphics[width=\linewidth, frame]{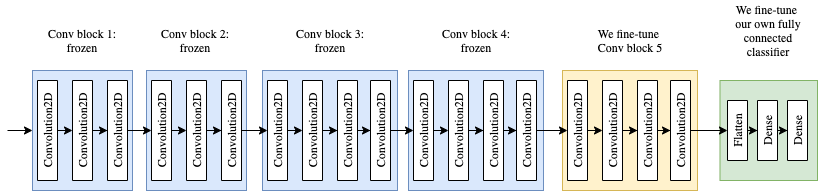}
	\caption{Fine Tuning of Transfer Learning}
 \label{mylabel9}
\end{figure}

Fine-tuning involves changing some of the convolutional layers by learning new weights. In Fig. ~\ref{mylabel9},  we have a network divided into three regions. The yellow region is a pre-trained model. The green region represents our dense layers for which we need to learn the weights. During training using a library such as Keras \cite{ketkar2017deep} and Tensorflow \cite{abadi2016tensorflow}, we can select certain layers and freeze the weights of those layers.

For example, we can select convolutional block one and then freeze all the weights of the convolutional layers, in this block only. This means that during training, everything else will change, but the weights of the convolutional layers in this block will not change. Similarly, we can keep frozen the convolutional layers of the next block as well as blocks three and four if we so wish. Then, we can fine-tune the convolutional layers that are closer to the dense layer. As a result, the initial layers of representation are kept constant, but new representations are learned by later layers (yellow region) as their weights change, evolve and get updated.

Thus, fine-tuning means unfreezing a few of the top layers of a frozen model base used for feature extraction. What we simply do is jointly train the newly added top part of the model (green region) consisting of dense layers, and the top convolutional layers (yellow region), for which we have unfrozen the weights.

Why fine-tune in this manner?
Because, we slightly adjust the more abstract representations of the model being reused to make them more relevant for the problem at hand. \citet{SUDHAKAR2021334} redesigned ResNet50 \cite{he2016deep} by changing the last layer with a fully connected dense layer to detect unknown malware samples without feature engineering. \citet{go9185490} proposed a visualization approach to classify the malware families by using a ResNeXt50 pre-trained model. The model achieved 98.86\% accuracy on the Malimg dataset \cite{nataraj2011malware}. \citet{CAYIR2021102133} built an ensemble pre-trained capsule network (CapsNet) \cite{SabourSara} based on the bootstrap aggregating approach. The model was trained and tested on two public datasets, Malimg, and BIG2015. Their model achieved F-Score 96.6\% on the Malimg dataset \cite{nataraj2011malware} and 98.20\% on the BIG2015 dataset\footnote{https://www.kaggle.com/c/malware-classification}. \citet{bensaoud2020classifying} used six convolutional neural network models for malware classification. Comparison among these models shows that the transfer learning model called Inception-V3 \cite{szegedy2016rethinking} achieved the current state-of-the-art in malware classification. \citet{khan2019analysis} evaluated ResNet and GoogleNet \cite{szegedy2015going} models for malware detection by converting an APK bytecode into grayscale image. Table 3 summarizes the most transfer learning models for malware classification. We conclude that CNN transfer learning models can be fine-tuned to specific image sizes that are robust enough and accurate to use malware image classification.

\begin{table*}
    \caption{Fine-tuned pre-trained models applied on different malware image datasets.}
    \centering
    \begin{tabular}{cccccccc}
        \toprule
        \multicolumn{4}{c}{\cellcolor[HTML]{FFFFFF}\textcolor[HTML]{333333}{Setting}} & \multicolumn{3}{c}{\cellcolor[HTML]{FFFFFF}\textcolor[HTML]{333333}{Average Accuracy}} & \cellcolor[HTML]{FFFFFF}\textcolor[HTML]{333333}{Our Dataset} \\
        \midrule
        \textcolor[HTML]{333333}{Pre-trained Model} & \textcolor[HTML]{333333}{Samples} & \textcolor[HTML]{333333}{Resize image} & \textcolor[HTML]{333333}{Epoch} & \textcolor[HTML]{333333}{Malimg} & \textcolor[HTML]{333333}{Microsoft Challenge} & \textcolor[HTML]{333333}{Drebin} & \textcolor[HTML]{333333}{Accuracy} \\
        \midrule
        EffNet B0 & 30,000 & 224 & 200 & 92.72\% & 90.45\% & 87.23\% & 94.59\% \\
        EffNet B1 & 30,000 & 240 & 200 & 95.64\% & 93.65\% & 88.91\% & 95.89\% \\
        EffNet B2 & 20,000 & 260 & 200 & 93.84\% & 91.78\% & 86.82\% & 94.12\% \\
        EffNet B3 & 15,000 & 300 & 400 & 90.32\% & 94.19\% & 89.35\% & 95.73\% \\
        EffNet B4 & 20,000 & 380 & 400 & 95.63\% & 96.68\% & 90.59\% & 97.98\% \\
        EffNet B5 & 25,000 & 456 & 400 & 80.19\% & 87.54\% & 84.23\% & 94.68\% \\
        EffNet B6 & 40,000 & 528 & 400 & 85.67\% & 83.82\% & 85.43\% & 93.54\% \\
        EffNet B7 & 30,000 & 600 & 1000 & 82.76\% & 80.76\% & 90.57\% & 88.45\% \\
        Inception V4 & 20,000 & 229 & 300 & 95.98\% & 93.21\% & 88.93\% & 96.39\% \\
        Xception & 20,000 & 229 & 200 & 89.50\% & 90.84\% & 84.39\% & 93.53\% \\
        CapsNet & 3,000 & 256 & 100 & 88.64\% & 72.69\% & 78.68\% & 92.65\% \\
        \bottomrule
    \end{tabular}
\end{table*}

\begin{figure*} [ht]
	\includegraphics[width=\linewidth, frame]{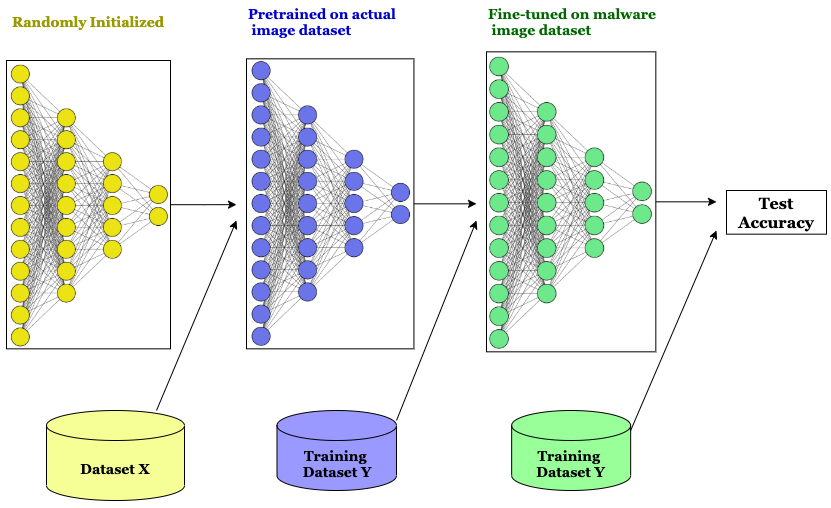}
	\caption{Transfer Learning steps}
 \label{mylabel10}
\end{figure*}

Fig. ~\ref{mylabel10} shows how to train the model on an image dataset. We randomly initialize the model, and then train the model on dataset X, which is a large-scale image dataset. This is the pre-training step. Next, we train the model on dataset Y; this dataset is typically smaller than dataset X. This is the fine-tuning step. 

State-of-the-art transfer learning models we have trained and evaluated for malware classification are EffNet B0, B1, B2, B3, B4, B5, B6, and B7 \cite{tan2019efficientnet}; Inception-V4 \cite{szegedy2017inception}, Xception \cite{chollet2017xception}, and CapsNet \cite{sabour2017dynamic} as shown in Table 3. The datasets used are our RGB malware image dataset and two other datasets, namely Malimg Dataset \cite{nataraj2011malware} and Microsoft Malware Dataset \cite{gibert2020rise}. The accuracy and loss curve plots for EffNet B1, B2, B3, B4, B5, B6, and B7 are shown in Appendix B and EffNet B0 shows in Fig. ~\ref{mylabel11}.
\begin{figure} [ht]
	\includegraphics[width=\linewidth, frame]{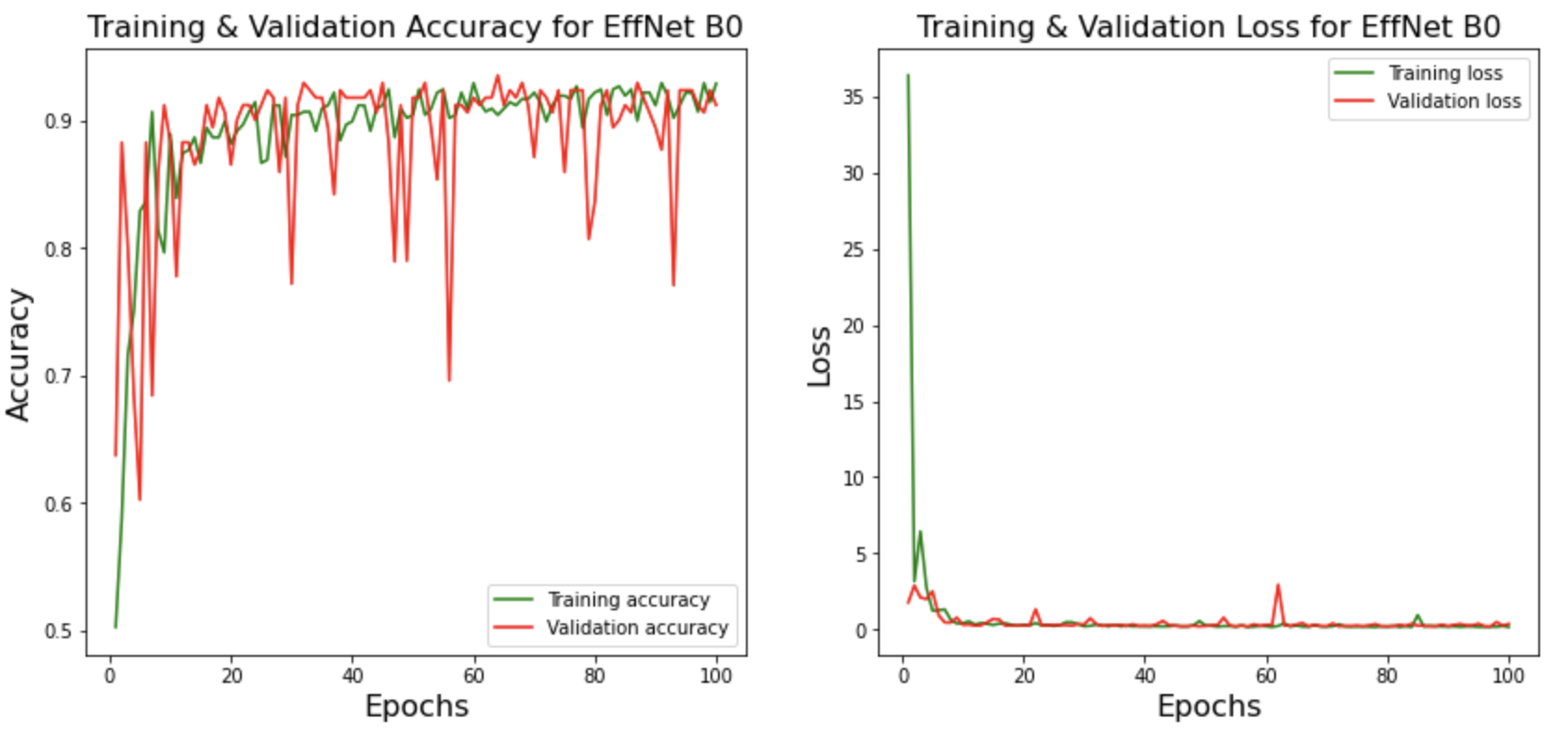}
	\caption{Training and testing for accuracy and loss of EfficientnetB0}
 \label{mylabel11}
\end{figure}

We found that the Inception-V4 model is most effective in classifying malware images among the ten models. In addition, the training times for each model increases with increase in the size of input images since the number of network cells grows quickly in GPU RAM.  
\subsection{Analysis of Transfer Learning for Malware Classification}
We found that transfer learning based image classification, with a small number of parameters to retrain successfully to classify malware images. On the other hand, we argue that scaled up wider and deeper transfer models with more parameters builds a new model that may improve performance. Inception-V3 and Inception-V4 for malware detection and classification avoid the inefficiencies in classifying unknown malware grayscale and RGB images among transfer learning classification model. There are many transfer learning models techniques such as batch normalization \cite{Kocaman9412907}, skip connections \cite{alaraimi2021transfer} that are designed to help in training, but the accuracy still needs to improve. For instance, ResNet-101 and ResNet-50 have similar accuracies in terms of malware detection even though they have very different deep networks \cite{eum2018going}. 

%[width=.4\textwidth,cols=2,pos=h]

%Comprehensive analysis of malware image classification techniques.
%Table 1: Machine Learning-Based Malware Detection Frameworks Comparison
\section{Natural Language Processing for Malware Classification}
Natural Language Processing (NLP) extracts valuable information so that a program is able to read, understand and derive meaning from human language text or speech. Malware data contain executable files, Microsoft Word files, macro files, logs from different operating systems, emails, network activities, etc. Many of these files contain extensive amounts of text; some others contain snippets of text mixed with code and other information. NLP can be used to enhance malware classification due to the extensive use of text or text-like content within malware. A critical requirement for malware text classification is using effective text representation in the form text encoding. The initial step in text encoding is preprocessing by removing a redundant opcode or API fragments, discarding unnecessary text.
 After tokenization, there are different types of non-sequential text representations \cite{jurafsky_martin} such as Bag of Words (BoW), Term Frequency Inverse document frequency matrices (TFIDF), Term document matrices (TDM), n-grams, One hot encoding, ASCII representations, and modren word embedding such as Word2vec \cite{mikolov2013efficient} and Sent2vec \cite{pagliardini2017unsupervised}. Table 4 presents text representation methods used in malware classification. Current word embeddings, when used in malware classification, do not carry much semantic and contextual significance. \citet{bensaoud2024cnn} proposed a novel model for malware classification using API calls and opcodes, incorporating a combined Convolutional Neural Network and Long Short-Term Memory architecture. By transforming features into N-gram sequences and experimenting with various deep learning architectures, including Swin-T and Sequencer2D-L, the method achieves a high accuracy of 99.91\%, surpassing state-of-the-art performance. \citet{mimura2021applying} designed NLP-based malware detection by using printable ASCII strings. The model can detect effectively packed malware and anti-debugging. Sequence to Sequence neural models are commonly used for natural languages processing and therefore used for malware detection as well.

\begin{table}
    \caption{The steps of encoding the domain by NLP.}
    \centering
    \begin{tabular}{lc}
        \toprule
        \textbf{Domain} & \textbf{Notes} \\
        \midrule
        www.uccs.edu & Start with domain \\
        uccs & Extract second level \\
        {[}"u","c","c","s"{]}  & Convert to sequence \\
        {[}21,3,3,18{]}  & \begin{tabular}[c]{@{}l@{}}Translate character to \\ numeric values\end{tabular}  \\
        {[}0,0,0,....,0,21,3,3,18{]} & Pad sequence \\
        \bottomrule
    \end{tabular}
\end{table}

\subsection{Sequence to Sequence Neural Models}
\begin{figure} [ht]
	\includegraphics[width=\linewidth, frame]{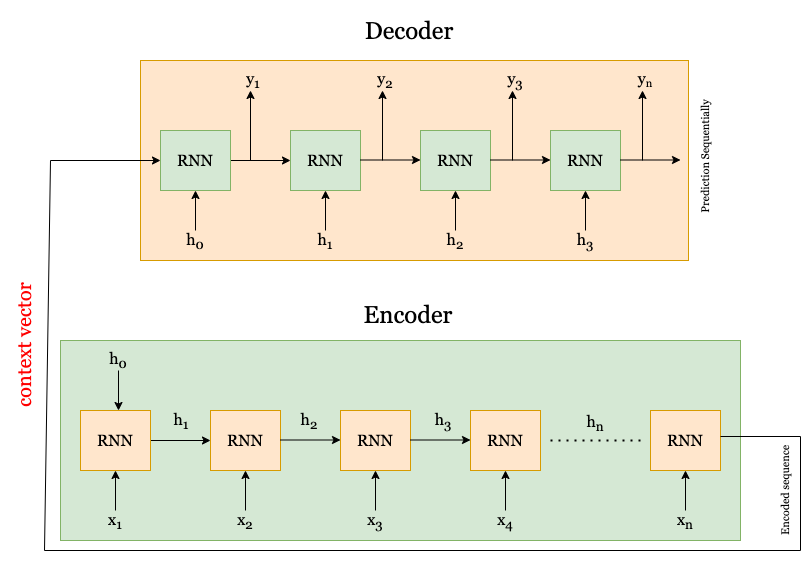}
	\caption{Encoder and decoder}
 \label{mylabel12}
\end{figure}

\begin{figure} [ht]
	\includegraphics[width=\linewidth, frame]{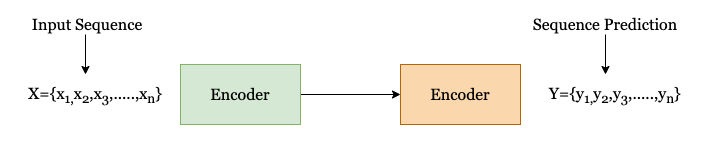}
	\caption{Encoder and decoder include RNNs}
  \label{mylabel13}
\end{figure}

\begin{figure} [ht]
	\includegraphics[width=\linewidth, frame]{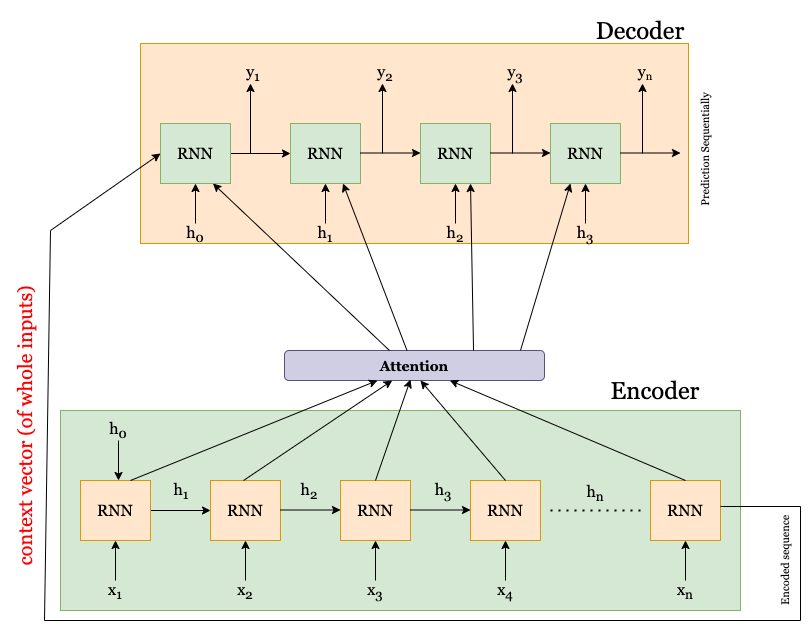}
	\caption{Encoder and decoder include RNNs with attention mechanism}
 \label{mylabel14}
\end{figure}

Attention mechanism \cite{luong2015effective} has achieved high performance in sequential learning applications such as machine translation \cite{Lu9576577}, image recognition \cite{Gao9391092}, text summarization \cite{almazrouei2021feasibility}, and text classification \cite{NIU202148}. Attention mechanism was designed to improve the performance of the encoder-decoder machine translation approach \cite{ren2021semface}. The encoder and decoder are usually many stacked RNN layers such as LSTM as shown in Fig. ~\ref{mylabel12}. The encoder converts the text into a fixed-length vector while the decoder generates the translation text from this vector. The sequence \{$x_1,x_2,..., x_n$\} can either be a  representation of text or image as shown in Fig. ~\ref{mylabel13}. In case of sequences, Recurrent Neural Networks (RNNs) can take two sequences with the same or arbitrary lengths. In Fig. ~\ref{mylabel14}, the encoder creates a compressed representation called context vector of the input, while the decoder gets the context vector to generate the output sequence. In this approach, the network is incapable of remembering dependencies in long sentences. This is because the context vector needs to handle potentially long sentences, and a shoot overall representation does not have the especially to store many potential dependencies.

\textbf{Attention in encoder-decoder:} 
\citet{bahdanau2014neural} proposed an encoder-decoder attention mechanism framework for machine translation. A single fixed context vector is created by an RNN by encoding the input sequence. Rather than using just the fixed vector, we can also use each state of the encoder along with the current decoder state to generate a dynamic context vector. There are two benefits; the first benefit is encoding information contained in a sequence of vectors not just in one single context vector. The second benefit is to choose a subset of these vectors adaptively while decoding the translation.

An attention mechanism is another Lego block that can be used in any deep learning model. \citet{vaswani2017attention} showed that an attention mechanism is apparently the only Lego block one needs. It improved the performance of a language translation model by dynamically choosing important parts of the input sequence that matter at a certain point in the output sequence. We can entirely replace traditional Recurrent Neural Network (RRN) blocks by an attention mechanism block. When dealing with sequential data, the attention mechanisms allow models to not only perform better but also train faster.

\begin{figure} [ht]
	\includegraphics[width=\linewidth, frame]{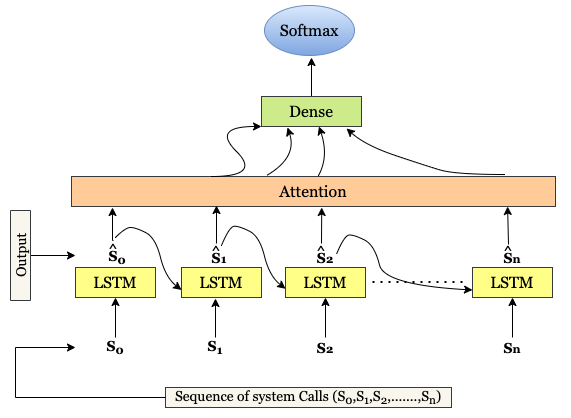}
	\caption{LSTM with attention mechanism for malware classification}
 \label{mylabel15}
\end{figure}

\textbf{Applying attention mechanism in malware classification:} \citet{or2021pay} added an attention mechanism to an LSTM model, which improved accuracy in malware classification. \citet{YAKURA2019101592} proposed a method by using Convolutional Neural Network with Attention Mechanism for malware image classification. \citet{mimura2020using} proposed a sliding local attention mechanism model (SLAM) based on API execution sequence. \citet{Ma8876839} proposed a malware classification framework (ACNN) based on two sections within the malware text, the assembly code and binary code, and converted them into multi-dimensional features. A CNN with attention mechanism for classification has a higher malware image classification accuracy than conventional methods \cite{YAKURA2019101592}. To build predictive models using LSTM and attention mechanism for malware classification, we need to add an embedding layer followed by an LSTM layer and dense layers . This approach is superior to capturing a long sequence of Windows API call sequences and using them directly \cite{Girinoto9354301} (see Fig. ~\ref{mylabel15}). Malware’s longer sequence can be addressed by attention mechanisms that can help detect short repeating patterns and other dependencies \cite{Agrawal8682899}. While attention mechanism improves accuracy, it suffers from the heavy computation.

\begin{table*}
    \caption{State-of-the-art deep learning models.}
    \centering
    \begin{tabular}{ccccc}
        \toprule
        \textbf{Ref} & \textbf{Deep Learning Approach} & \textbf{OS} & \textbf{Features} & \textbf{Accuracy} \\
        \midrule
        \citet{kim2022mapas} & MAPAS & Android & API call graphs & 91.27\% \\
        \citet{onwuzurike2019mamadroid} & MaMaDroid & Android & API calls & 84.99\% \\
        \citet{KIM2022102501} & Deep Generative Model & Android & \begin{tabular}[c]{@{}c@{}}Dalvik code, \\ API call, \\ Malware images,\\ developers’ signature\end{tabular} & 97.47\% \\
        \citet{Olani9770351} & DeepWare & Windows/Linux & HPC & 96.8\% \\
        \citet{Lian9722781} & Multi-Modal Deep Learning & Windows & \begin{tabular}[c]{@{}c@{}}Grayscale image, \\ Byte/Entropy \\ Histogram\end{tabular} & 97.01\% \\
        \citet{BENSAOUD2022103057} & \begin{tabular}[c]{@{}c@{}}Deep multi-task \\ learning\end{tabular}  & \begin{tabular}[c]{@{}c@{}}Windows\\ Android\\ Linux\\ MacOS\end{tabular} & \begin{tabular}[c]{@{}c@{}}Grayscale\\ color image\end{tabular} & 99.97\% \\
        \bottomrule
    \end{tabular}
\end{table*}

Table 5 shows various approaches and their corresponding accuracies. The methods presented, including MAPAS, MaMaDroid, Deep Generative Model, DeepWare, Multi-Modal Deep Learning, and Deep Multi-Task Learning, employ diverse techniques such as API call graph analysis, static analysis, and hybrid deep generative models. Particularly, these methods are evaluated on distinct datasets, indicating that the comparisons are not based on the same dataset. The authors aim to convey the effectiveness of these models in detecting malware across different datasets and scenarios. However, a comprehensive overview of the comparative performance of these methods is needed, highlighting their strengths and capabilities in addressing the challenges of malware detection.
\section{Deep Learning for Cryptographic Ransomware} 
Cryptography has been used traditionally for military and government use, to keep secrets from the enemy. Today most of us use cryptography when we use commercial websites or services. For example, we use it to protect our emails. A lot of countries try to control the export of cryptography to make sure that good cryptographic algorithms are not in the hands of criminals, enemies, or adversaries. This is the idea behind export administration, and  regulations as codified in International Traffic in Arms Regulations (ITAR)\footnote{https://csrc.nist.gov/glossary/term/itar}. In addition, we have various agreements like the Wassenaar Arrangement\footnote{https://www.federalregister.gov/documents/2022/08/15/2022-17125/implementation-of-certain-2021-wassenaar-arrangement-decisions-on-four-section-1758-technologies}, where a number of countries got together and developed an agreement for what cryptographic  elements can be exported and imported without any type of restrictions. This agreement allows publicly available cryptographic algorithms to be distributed freely. Cryptography provides various security capabilities for us. 
\begin{itemize}
\item \textbf{Confidentiality:} 
To protect our intellectual property from somebody else being able to get hold of it.
\item \textbf{Non-repudiation:} 
To repudiate is to deny. For example, if we use digital signatures, we can provide proof that the message came from the person who signed. We can link the signed document to a trusted person, which gives us trust or assurance in the world of e-contracts and e-commerce. The signer cannot repudiate or deny being the source of the document. 
\item \textbf{Integrity:} 
Hashing provides integrity, to know that a message was not changed either accidentally or intentionally as it was transmitted or stored. Integrity is built into implementation of electronic communication services today using such as SHA algorithms\footnote{https://csrc.nist.gov/glossary/term/sha} and MD5\footnote{https://csrc.nist.gov/glossary/term/md5}.

\item \textbf{Proof-of-Origin:} 
Cryptography can be used to prove where a message came from, the idea of Proof-of-Origin.

\item \textbf{Authenticity:} 
The idea is to ensure that communication is with the intended person. For example, if we go to a bank’s website, then we want to be sure that the website is truly of that bank, not that of an impostor or somebody else masquerading as that bank. 
\end{itemize}

\subsection{Operations of Cryptography}
Cryptographic algorithms come in three basic flavors: Symmetric, Asymmetric, and Hash algorithms. Each of these different types of algorithms serves a different purpose, but all work together in a cryptography system. 

Cryptography is a key to keeping communicated information secret by converting it into an unreadable code that is hard to break. To encrypt or encipher is to take a plaintext message and convert it into something unreadable to anyone who does not have a key. To decrypt or decipher is the reverse step.

\begin{figure} [ht]
\centering
	\includegraphics[width=\linewidth, frame]{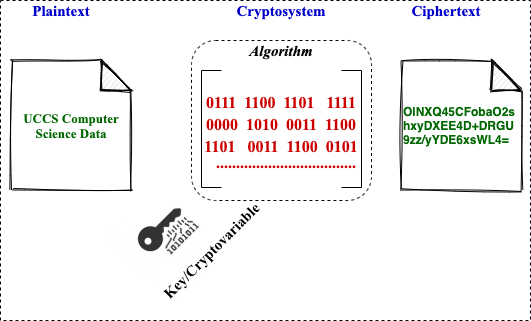}
	\caption{Crypto Action}
 \label{mylabel16}
\end{figure}

In Fig. ~\ref{mylabel16}, the basic action includes plaintext being fed into a cryptosystem. This process is used to encrypt and decrypt a message. It contains an algorithm that uses a mathematical process to convert a message from plaintext to ciphertext and then back again. The algorithm includes a key or a cryptovariable. The variable is used by the algorithm during the encryption and decryption processes. Typically the key is a secret password, passphrase, or PIN chosen either by the person or by the tool that encrypts the message. This combination of the key (or a cryptovariable) and the algorithm in the cryptosystem produces a unique ciphertext.

In the symmetric algorithm family, a symmetric key is one that is a shared secret between the sender and receiver of the information. The same key used for encryption is also used for decryption. It is not safe to send a copy of the key along with the message that it encrypts. We need to use another mode of communication to transmit the key. For example, Ahmed sends the symmetric key to Bryan using a certain secure node of communication. Once Bryan has the key, Ahmed can encrypt the plaintext message into ciphertext and send it over a public network to Bryan with confidence that it will remain encrypted until Bryan decides to decrypt with the received key. 

{\small Multiple attacks, such as a man-in-middle attack, brute force attack, biclique attack, ciphertext only attack, known plaintext attack, chosen plaintext attack, chosen ciphertext attack, and chosen text attack can discover the key to find the plaintext. Attackers know the mathematical relationship of the keys for some algorithms, such as Advanced Encryption Standard (AES) \cite{heron2009advanced}, Triple DES \cite{sasi2015survey}, Blowfish \cite{mahendra2022classification}, and Rivest-Shamir-Adleman (RSA) \cite{kota2022}. We perform cryptanalysts using statistical measures to try to get the cipher type, but a cryptanalyst can only test as many solvers via trial and error to test if the ciphertext was encrypted using a specific cipher. Machine learning can tell us what type a cipher is \cite{Lee9551956}. The cipher type detection problem is a classification problem. We can use statistical values as features for machine learning.}

\subsection{Connection Between Deep Learning and Cryptography}

\begin{figure*}[ht]
\centering
\includegraphics[height=100mm,width=160mm]{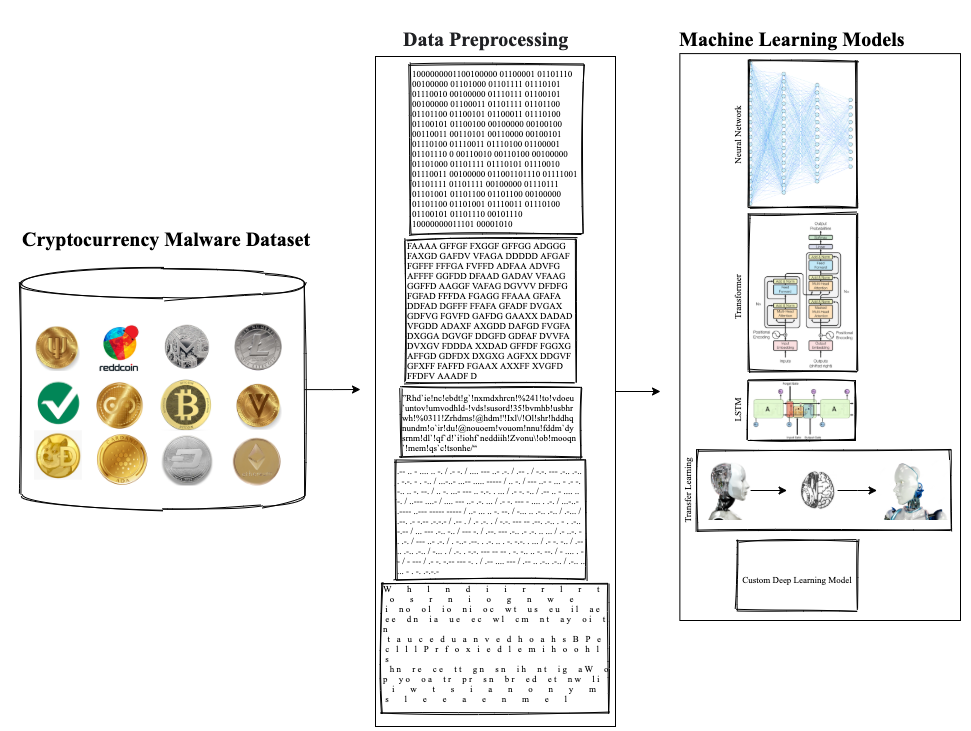}
  \caption{Cryptocurrency malware detection using machine learning}
   \label{mylabel17}
\end{figure*}

\begin{figure*} [ht]
	\includegraphics[width=\linewidth, frame]{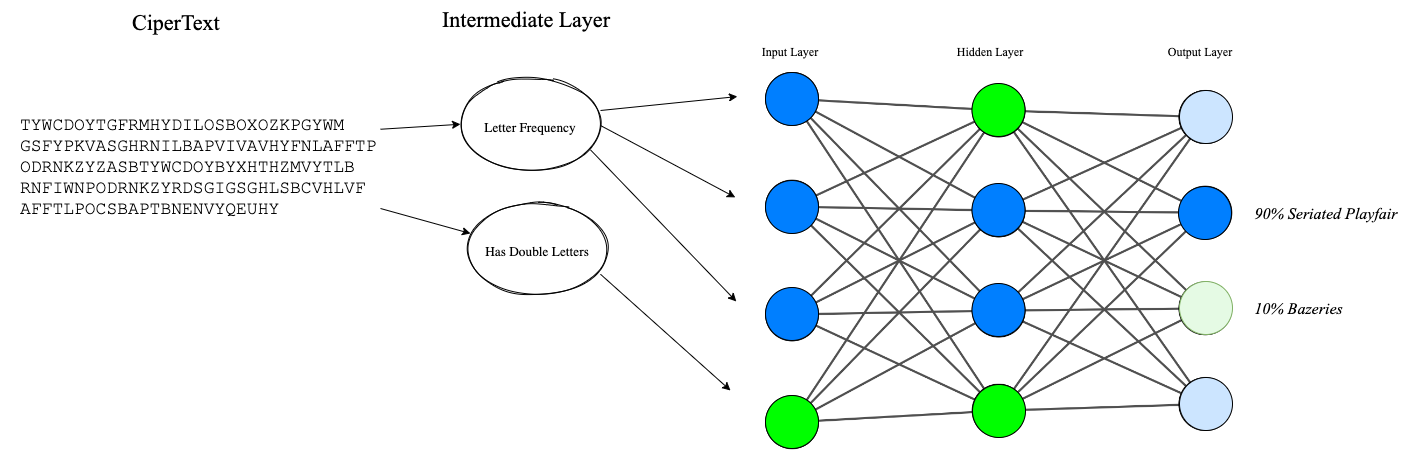}
	\caption{Detect the Cipher Type With Neural Networks}
  \label{mylabel18}
\end{figure*}

A neural network can deal with the complexity of computation applied to perform cryptography. Instead of giving an image to a neural network, we can give ciphertext to the neural network to classify the kind of algorithm that was used to obtain the ciphertext, as shown in Fig. ~\ref{mylabel17}. To build a machine learning model, we can represent different features of the cipher, which cryptanalysts usually use to identify them. We need to put an intermediate layer between the network and ciphertext that computes the features, such as Unigram frequencies, Bigram frequencies, Index of Coincidence IoC, HasDoubleLetters, etc., and then we can train the network with millions of ciphertext and all American Cryptogram Association (ACA) cipher types. For example, in Fig. ~\ref{mylabel18}, the three blue neural networks are given the frequencies of N-grams (1-grams, 2-grams, 3-grams, 4-grams, etc.), and the green neural network computes HasDoubleLetters. Then we have a hidden layer that connects the input and output layers. Finally, in this case the designed neural network shows 90\% Seriated Playfair, and the green neural network shows 10\% Bazeries. \citet{baksi2022machine} designed a machine-learning model for differential attacks on the non-Markov 8-round GIMLI cipher and GIMLI hash. They applied multi-layer perceptron (MLP), Convolutional Neural Networks (CNN), and Long Short-Term Memory (LSTM).

The ransomware families to encrypt data and force the victim to make payment via cryptocurrency include WannaCry, Locky, Stop, CryptoJoker, CrypoWall, TeslaCrypt, Dharma, Locker, Cerber, and GandCrab. Recently, deep learning algorithms have been used for cryptography \cite{KOK2020102646}. \citet{ding2020deepedn} proposed DeepEDN to fulfill the process of encrypting and decrypting medical images. \citet{kim2021convolutional} proposed detection of cryptographic ransomware using Convolutional Neural Network. Their model prevents crypto-ransomware infection by detecting a block cipher algorithm. \citet{sharmeen2020avoiding} proposed an approach to extract the intrinsic attack characteristics of unlabeled ransomware samples using a deep learning-based unsupervised learned model. \citet{fischer2019stack} designed a tool to detect security vulnerabilities of cryptographic APIs in Android by achieving an average AUC-ROC of 99.2\%. 

\begin{figure*} [ht]
	\includegraphics[width=\linewidth, frame]{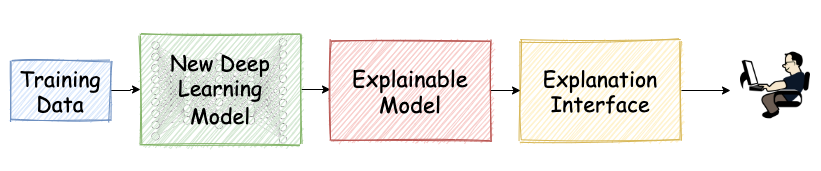}
	\caption{Using explainable artificial intelligence in deep learning }
 \label{mylabel19}
\end{figure*}

\section{Explainable Artificial Intelligence (XAI)}
Explainable Artificial Intelligence (XAI) is a rapidly emerging field that focuses on creating transparent and interpretable models (see Fig. ~\ref{mylabel19}). In the context of malware detection, XAI can help security experts and analysts understand how a machine learning model arrived at its decisions, making it easier to identify and understand false positives and false negatives. By applying XAI techniques, such as Local Interpretable Model-Agnostic Explanations (LIME) \cite{ribeiro2016should} or Deep Learning Important Features (DeepLIFT) \cite{shrikumar2017learning}, security teams can gain insights into the most important features and decision-making processes of the model. This can help them identify areas where the model may be vulnerable to evasion or identify new malware strains that the model may have missed. Ultimately, XAI can improve the trustworthiness and reliability of machine learning models for malware detection, enabling more effective threat detection and response. 

\citet{nadeem2022sok} provided a comprehensive survey and analysis of the current state of research on explainable machine learning (XAI) techniques for computer security applications. The paper highlights the challenges and opportunities for adopting XAI in the security domain and discusses several approaches for designing and evaluating explainable machine learning models. \citet{vivek2022explainable} proposed an approach for detecting ATM fraud using explainable artificial intelligence (XAI) and causal inference techniques. They presented a detailed analysis of the proposed method and highlighted its effectiveness in improving the accuracy and interpretability of ATM fraud detection systems. \citet{kinkead2021towards} proposed an approach that uses LIME to identify important locations in the opcode sequence that are deemed significant by the Convolutional Neural Network (CNN). \citet{mclaughlin2017deep} used LRP \cite{bach2015pixel} and DeepLift \cite{shrikumar2017learning} methods to identify the opcode sequences for most malware families, and they demonstrated that the CNN, while using the DAMD dataset, learned patterns from the underlying op-code representation. \citet{hooker2019benchmark} proposed a method to remove relevant features detected by an XAI approach and verify the accuracy degradation. \citet{lin2021you} presented seven different XAI methods and automated the evaluation of the correctness of explanation techniques. The first four XAI methods are white-box approaches to determine the importance of input features: Backpropagation (BP), Guided Backpropagation (GBP), Gradient-weighted Class Activation Mapping (GCAM), and Guided GCAM (GGCAM). The last three are black-box approaches that observe an essential feature in the output probability using perturbed samples of the input: Occlusion Sensitivity (OCC), Feature Ablation (FA), and Local Interpretable Model-Agnostic Explanations (LIME).

\citet{guo2018lemna} proposed an approach called Explaining Deep Learning based Security Applications (LEMNA) for security applications, which generates interpretable features to explain how input samples are classified. \citet{kuppa2020black} presented a comprehensive analysis of the vulnerability of XAI methods to adversarial attacks in the context of cybersecurity, discussing potential risks associated with deploying XAI models in real-world applications, and proposing a framework for designing robust and secure XAI systems. \citet{rao2021zero} proposed an approach to protect and analyze systems against the alarm-flooding problem using the NSL-KDD dataset. They included a Security Information and Event Management (SIEM) system to generate a zero-shot method for detecting alarm labels specific to adversarial attacks. Although explainable artificial intelligence (XAI) has gained significant attention, its effectiveness in malware detection still requires further investigation to fully comprehend its performance.

\section{Adversarial Attack on Deep Neural Networks}
Adversarial examples refer to maliciously crafted inputs to machine learning models designed to deceive the model into making incorrect predictions. Deep detection in this context refers to the use of deep learning models for detecting and classifying objects or patterns in the input data. Adversarial examples can be specifically crafted to evade deep detection models and cause them to misclassify or miss the target objects or patterns. Therefore, adversarial examples can be seen as a type of attack on deep detection models. Adversarial examples can be generated using a variety of techniques, including optimization-based approaches and perturbation-based approaches, and can be used for various objectives, including evasion attacks and poisoning attacks. \citet{zhong2023malfox} proposed a novel adversarial malware example generation method called Malfox, which uses conditional generative adversarial networks (conv-GANs) to generate camouflaged adversarial examples against black-box detectors. The presented method was evaluated on two real-world malware detection systems, and the results showed that Malfox achieved high attack success rates while maintaining low detection rates. \citet{zhao2023sage} proposed a new method called SAGE for steering the adversarial generation of examples with accelerations. The technique combines the advantages of gradient-based and gradient-free methods to generate more effective and efficient adversarial examples. 

The development of defense mechanisms against adversarial attacks is a computationally expensive process, which can potentially affect the performance of the deep learning model. In addition, adversarial examples can impact the generalization ability of deep learning models, resulting in poor performance on new and unseen data. Moreover, generating adversarial examples can be computationally intensive, especially for large datasets and complex models, which can hinder the practical deployment of deep learning models in real-world applications. Thus, further research is required to improve the efficiency and effectiveness of defense mechanisms, as well as the generalization ability and robustness of deep learning models to adversarial attacks.

\citet{hu2023generating} proposed a method to generate adversarial malware examples using Generative Adversarial Networks (GANs) for black-box attacks. Their results show that the generated adversarial malware samples can evade detection by existing machine learning models while maintaining high similarity to the original malware. \citet{ling2023adversarial} conducted a survey of the state-of-the-art in adversarial attacks against Windows PE malware detection, covering various types of attacks and defense mechanisms. The authors also provided insights on potential future research directions in this area. \citet{xu2023ofei} proposed a semi-black-box adversarial sample attack framework called Ofei that can generate adversarial samples against Android apps deployed on a DLAAS platform. The framework utilizes a multi-objective optimization algorithm to generate robust and stealthy adversarial samples. \citet{qiao2022adversarial} proposed an adversarial detection method for ELF malware using model interpretation and show that their method can effectively identify adversarial ELF malware with high accuracy. The proposed approach combines random forests and LIME to identify the most important features and thus improve the interpretability and robustness of the model. \citet{meenakshi2023optimised} proposed a defensive technique using Curvelet transform to recognize adversarial iris images, optimizing the image classification accuracy. The designed method was shown to be effective against several existing adversarial attacks on iris recognition systems. \citet{pintor2022indicators} introduced a method for debugging and improving the optimization of adversarial examples by identifying and analyzing the indicators of attack failure. The proposed method can help to improve the robustness of deep learning models against adversarial attacks.
\section{Conclusion}
Machine learning has started to gain the attention of malware detection researchers, notably in malware image classification and cipher cryptanalysis. However, more experimentation is required to understand the capabilities and limitations of deep learning when used to detect/classify malware. Deep learning can reduce the need for static and dynamic analysis and discover suspicious patterns. In the future, researchers may consider developing more accurate, robust, scalable, and efficient deep learning models for malware detection systems for various operating systems. Finally, multi-task learning and transfer learning can provide valuable results in classifying all types of malware. Furthermore, we show that the significant challenges of deep learning approaches that need to be considered are hyperparameters optimization, fine-tuning, and size and quality of datasets when features are overweighted or overrepresented. We also illustrate the opportunities and challenges of XAI in deep learning as well as future research directions in the context of malware detection. Finally, we presented the idea of adversarial attacks on deep neural networks by introducing small, carefully crafted perturbations to input data in order to cause misclassification or reduce model performance.

\bibliographystyle{model1-num-names.bst}
\medskip
\bibliography{cas-refs}

\section{Appendix A: File Types}
\textcolor{blue}{. tbk, .jpeg , . brd, .dot , .jpg , .rtf , .doc , .js , .sch , .3dm , .mp3 , .sh , .3ds , .key , .sldm , .3g2 , .lay , .sldm , .mkv , .std , .asp , .mml , .sti , .avi , .mov , .stw , .backup , . jsp, .suo , .bak , .mp4 , .svg , .bat , .mpeg , .swf , .bmp , .mpg , .sxc , .rb , .msg , .sxd , .bz2 , .myd , .sxi , .c , .myi , .sxm , .cgm , .nef , .sxw , .class , .odb , .tar , .cmd , .odg , .123 , . onetoc2, .odp , .tgz , .crt , .ods , .tif , .3gp , .lay6 , .sldx , .7z , .ldf , .slk , .vsd , .m3u , .sln , .aes , .m4u , .snt , .ai , .max , .sql , . ppam, .mdb , .sqlite3 , .asc , .mdf , .sqlitedb , .asf , .mid , .stc , .asm , .cs , .odt , .tiff , .csr , .cpp , .txt , .csv , .pas , .vmx , .docb , .pdf , .vob , .docm , .pem , . accdb, .docx , .pfx , .vsdx , .602 , . p12, .wav , .dotm , .pl , .wb2 , .dotx , .png , .wk1 , .dwg , .pot , . xltx, .edb , .potm , .wma , .eml , .potx , .wmv , .fla , .ARC , .xlc , .flv , .pps , .xlm , .frm , .ppsm , .xls , .gif , .ppsx , .xlsb , .gpg , .ppt , .xlsm , .gz , .pptm , .xlsx , .h , .pptx , .xlt , .hwp , .ps1 , .xltm , .ibd , .psd , .wks , .iso , .pst , .xlw , .jar , .rar , . djvu, .java , .raw., .ost , .uop , .db , .otg , .uot , .dbf , .otp , .vb , .dch , .ots , .vbs , .der” , .ott , .vcd , .dif , .php, .vdi , .dip , .PAQ , .vmdk , .zip}
\section{Appendix B: The Accuracy and Loss Curves Plots}
\begin{figure} [ht]
	\includegraphics[width=\linewidth, frame]{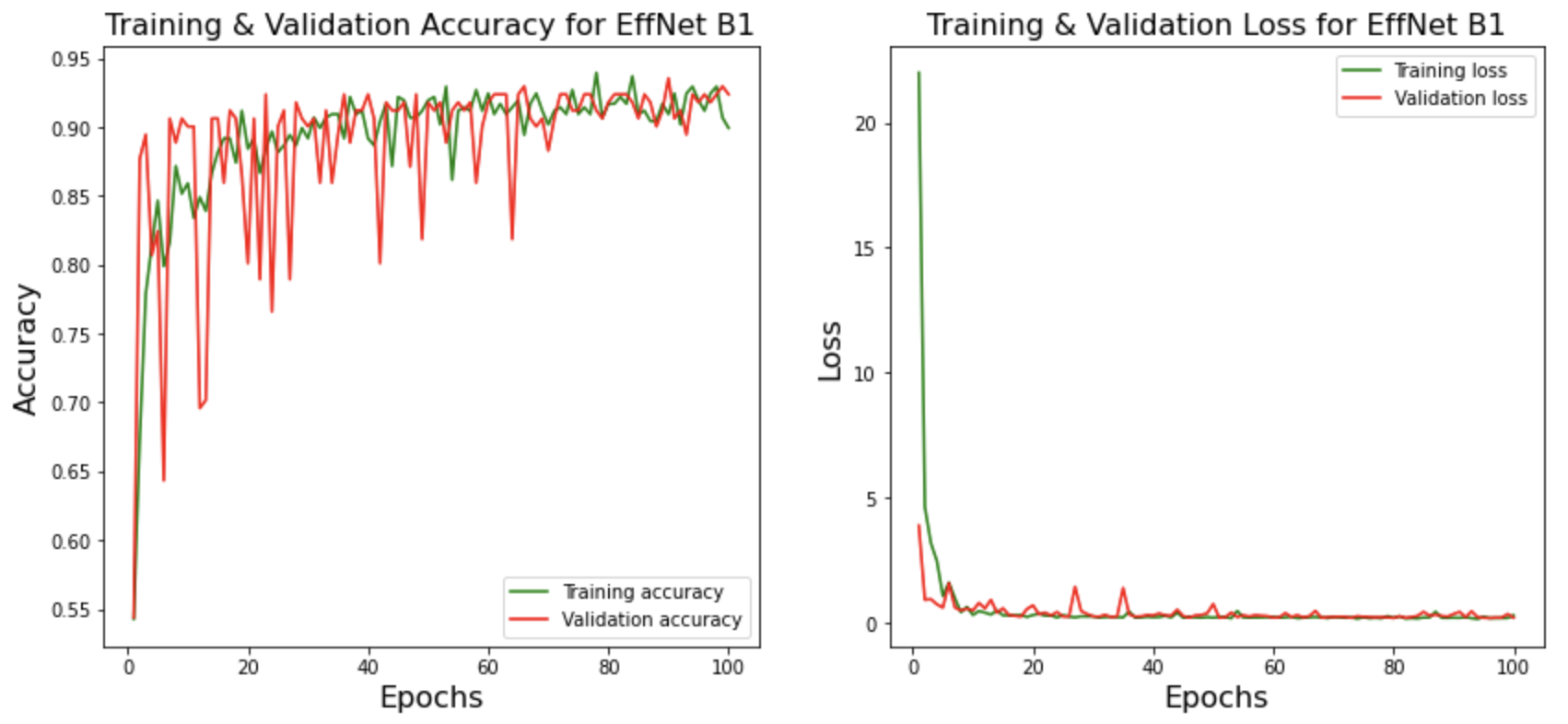}
	\caption{Training and testing for accuracy and loss of EfficientnetB1}
\end{figure}

\begin{figure} [ht]
	\includegraphics[width=\linewidth, frame]{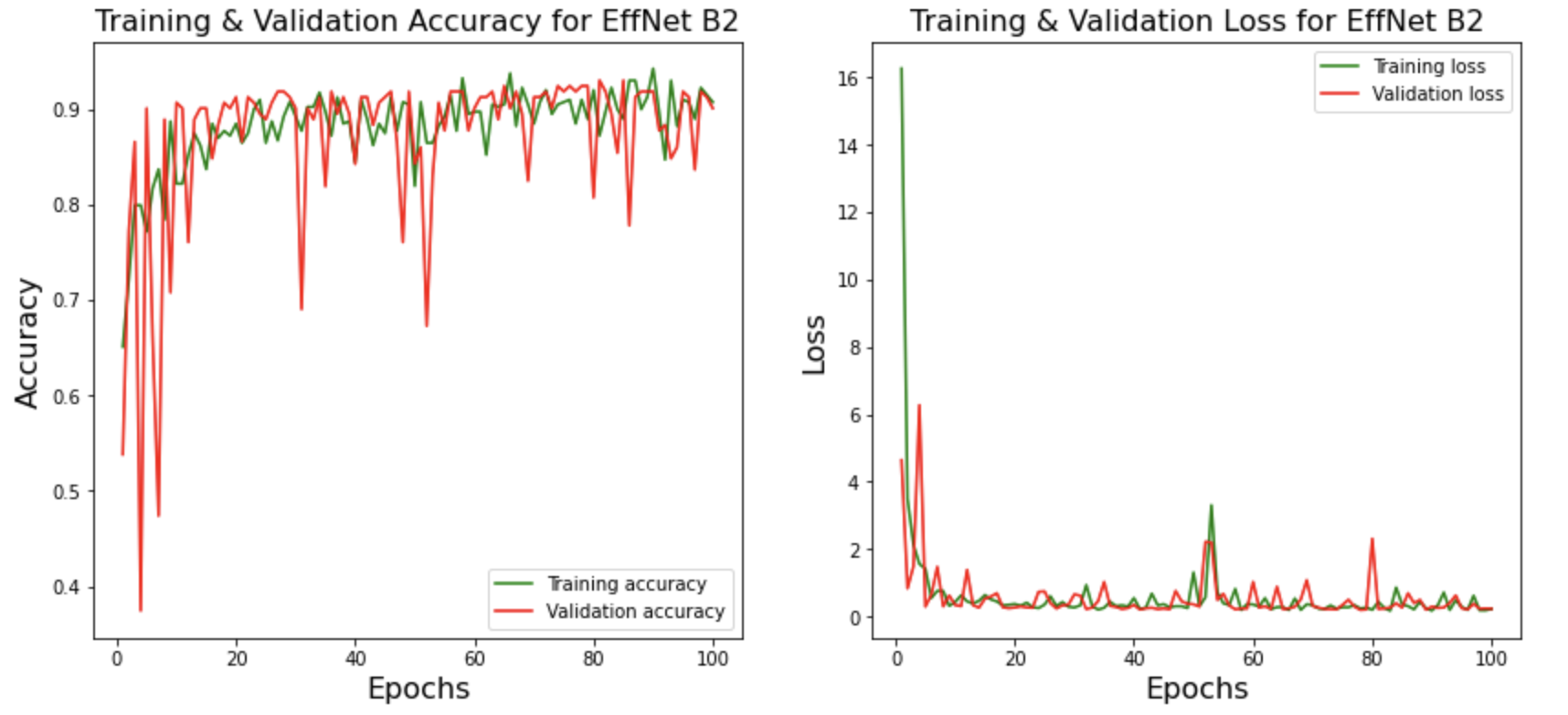}
	\caption{Training and testing for accuracy and loss of EfficientnetB2}
\end{figure}

\begin{figure} [ht]
	\includegraphics[width=\linewidth, frame]{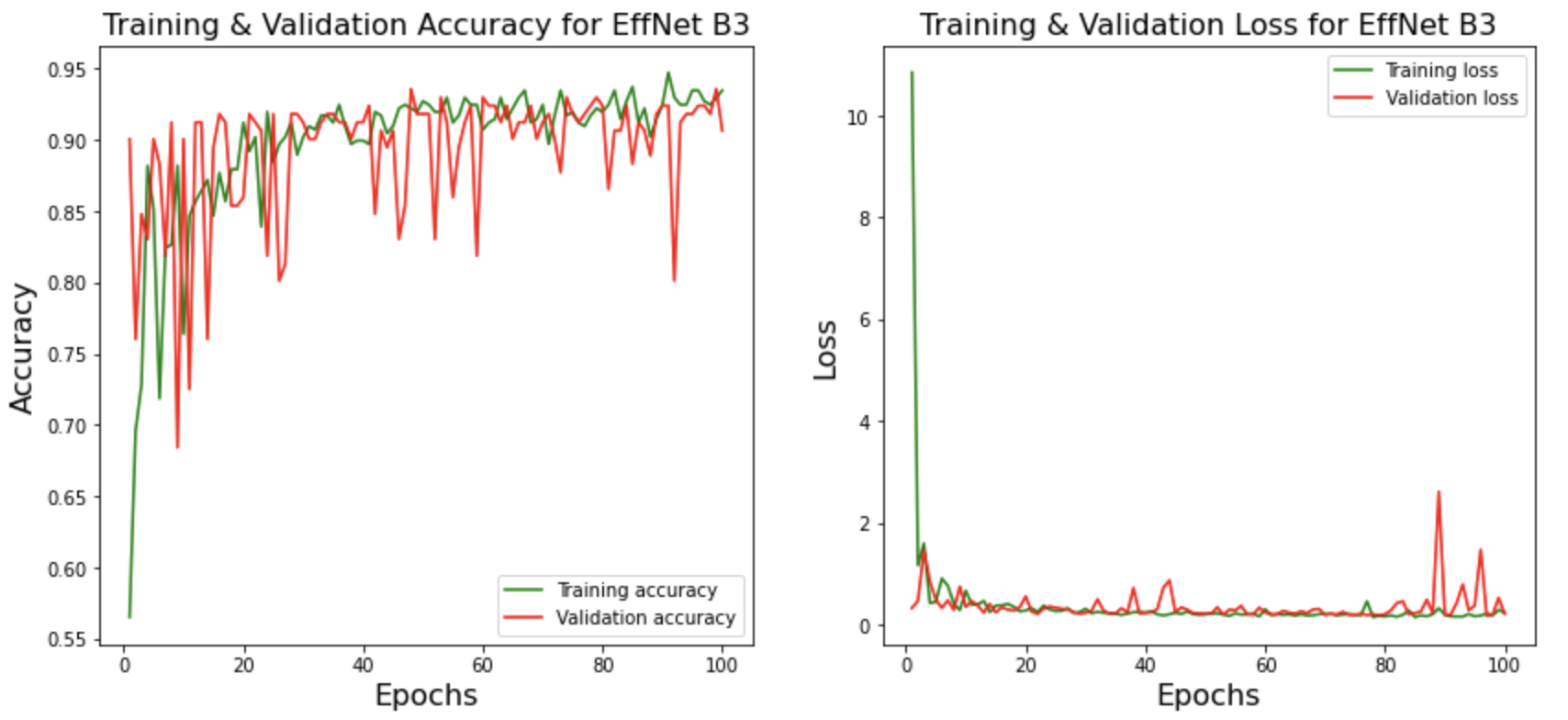}
	\caption{Training and testing for accuracy and loss of EfficientnetB3}
\end{figure}

\begin{figure} [ht]
	\includegraphics[width=\linewidth, frame]{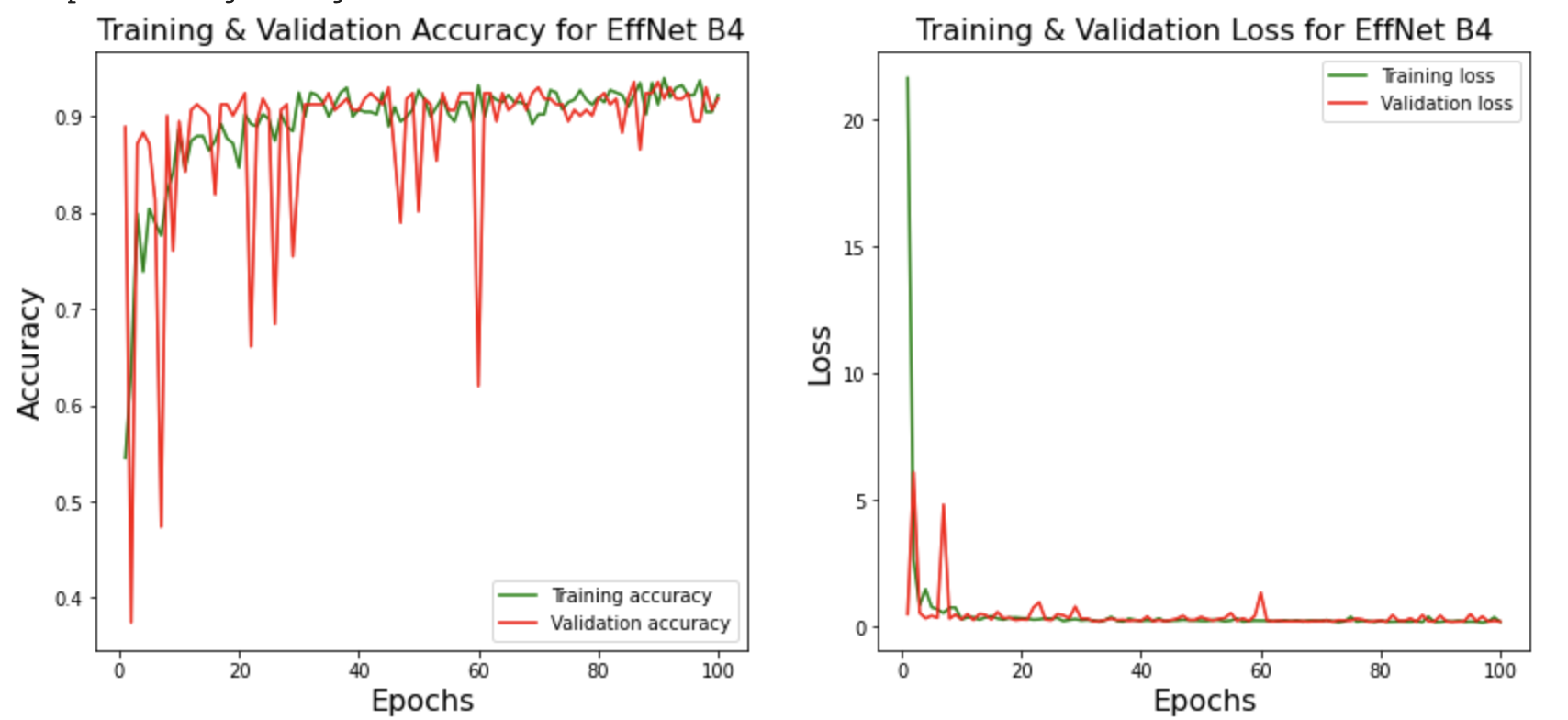}
	\caption{Training and testing for accuracy and loss of EfficientnetB4}
\end{figure}

\begin{figure} [ht]
	\includegraphics[width=\linewidth, frame]{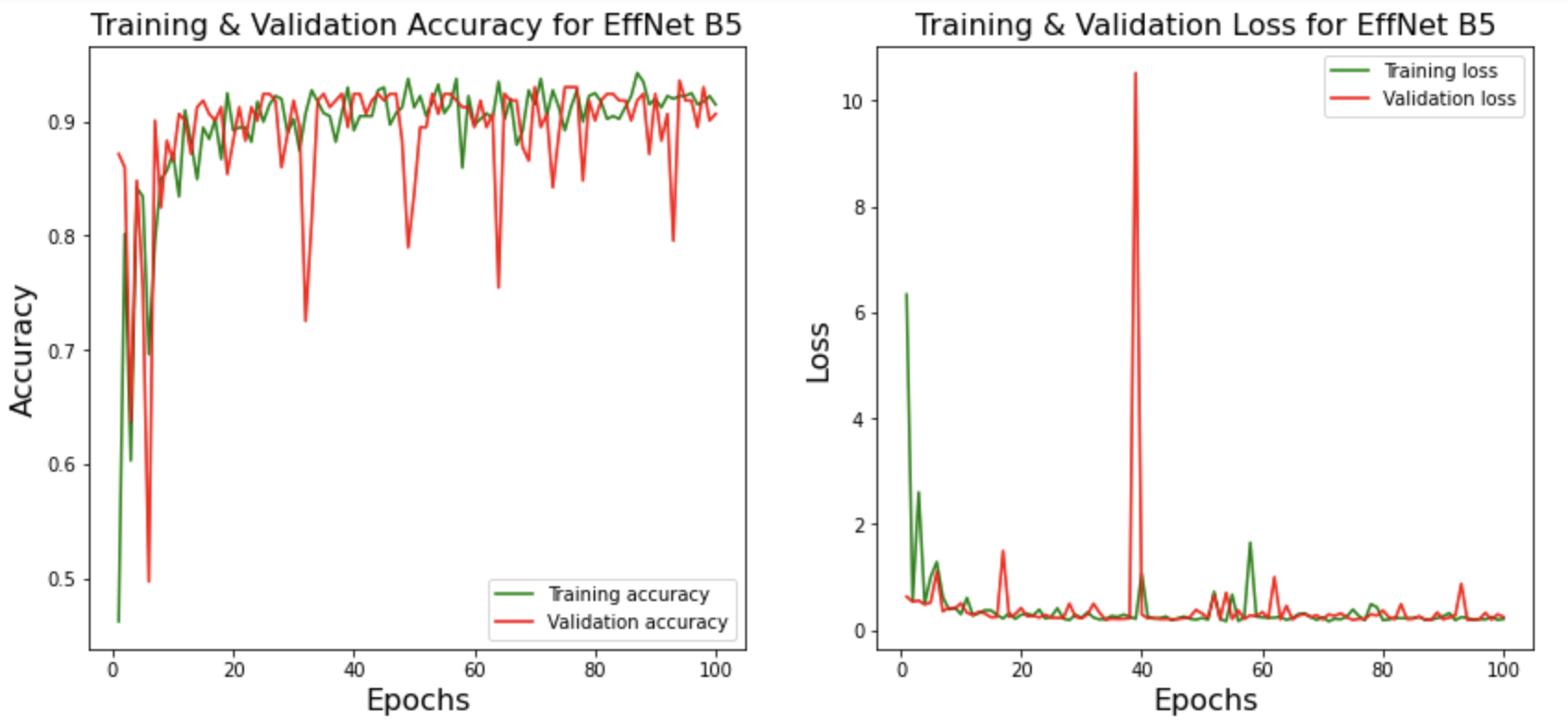}
	\caption{Training and testing for accuracy and loss of EfficientnetB5}
\end{figure}

\begin{figure} [ht]
	\includegraphics[width=\linewidth, frame]{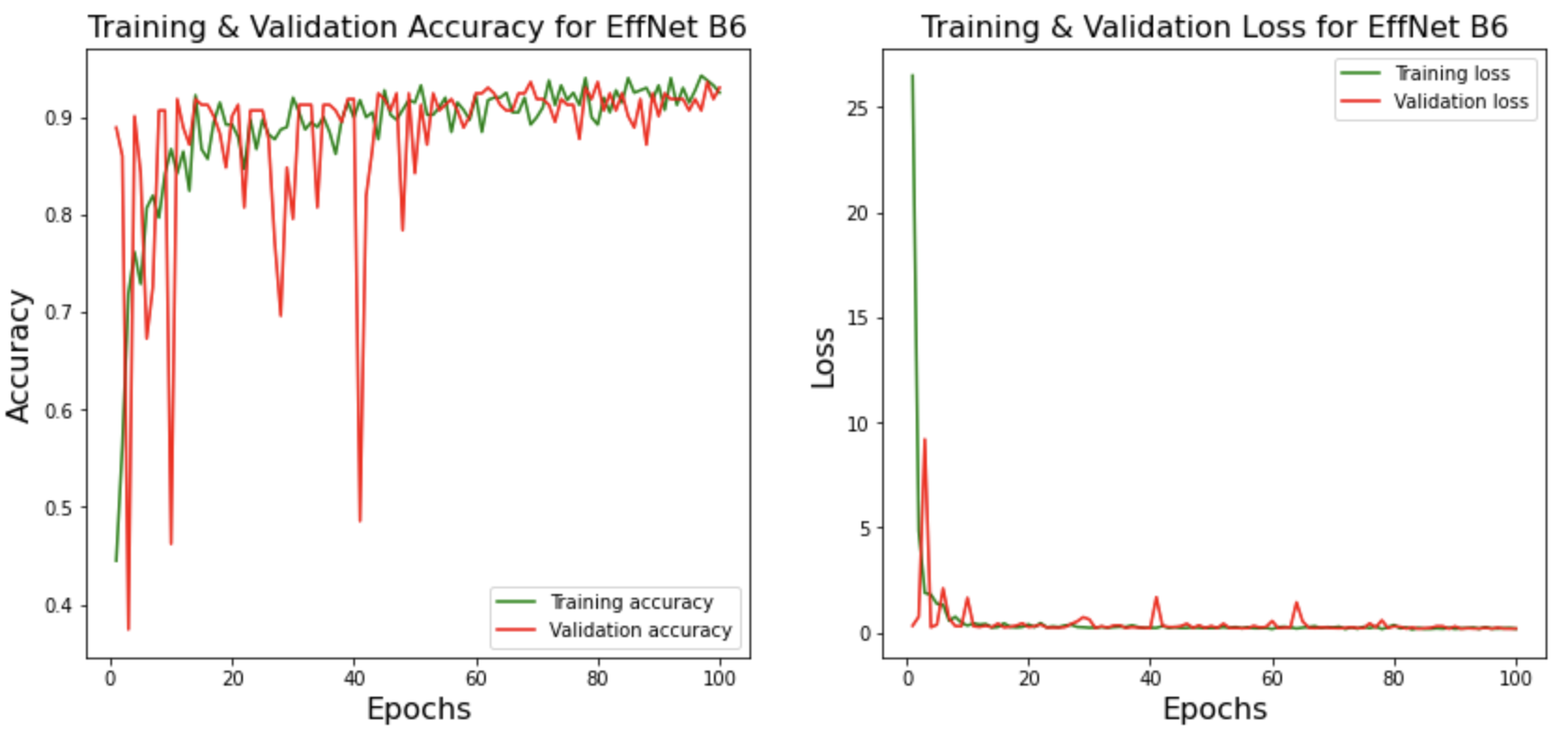}
	\caption{Training and testing for accuracy and loss of EfficientnetB6}
\end{figure}

\begin{figure} [ht]
	\includegraphics[width=\linewidth, frame]{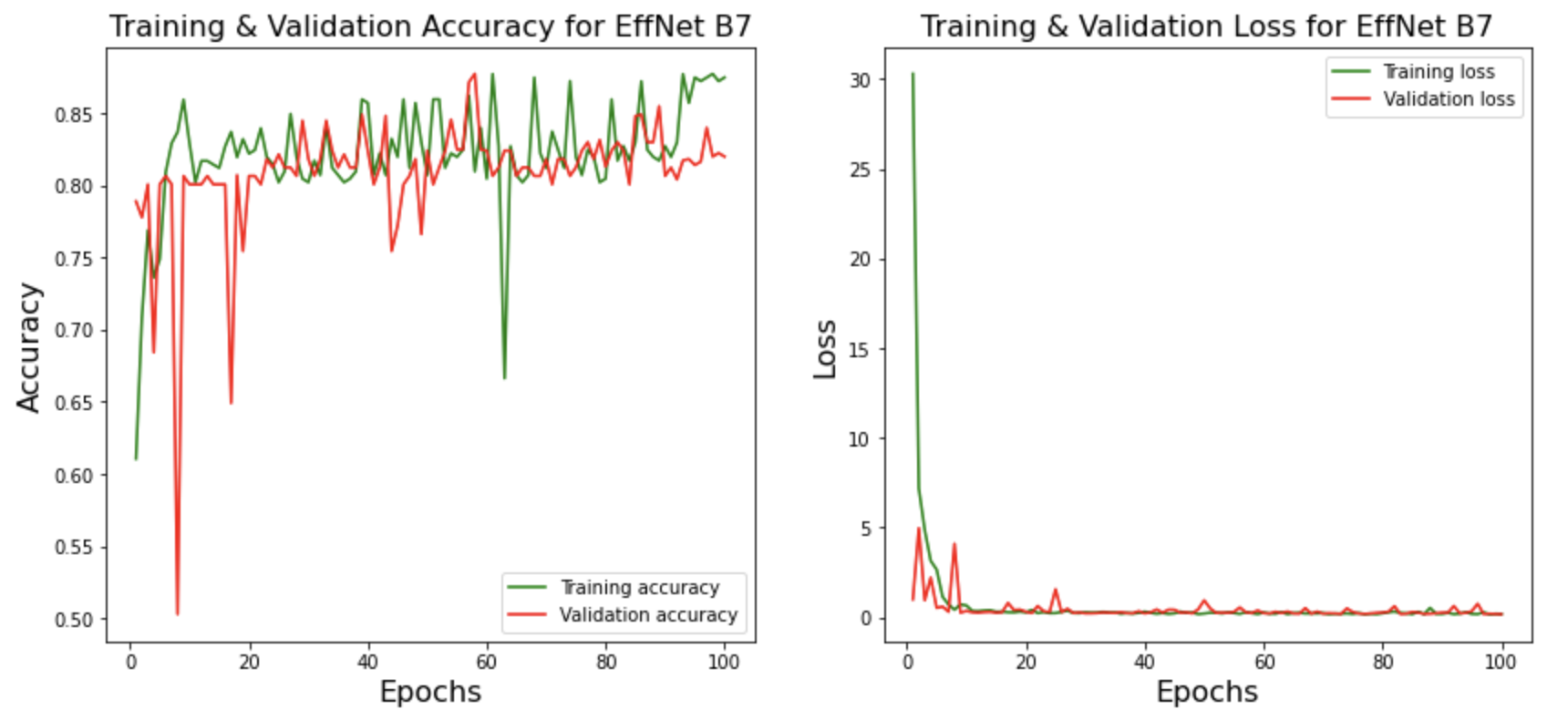}
	\caption{Training and testing for accuracy and loss of EfficientnetB7}
\end{figure}

\end{document}